\PassOptionsToPackage{table}{xcolor}

\documentclass[]{bytedance_seed}

\usepackage{amsmath,amsfonts,bm}

\def\eqref#1{equation~\ref{#1}}

\def\1{\bm{1}}

\DeclareMathAlphabet{\mathsfit}{\encodingdefault}{\sfdefault}{m}{sl}
\SetMathAlphabet{\mathsfit}{bold}{\encodingdefault}{\sfdefault}{bx}{n}

\usepackage{amsmath}
\usepackage{amssymb}
\usepackage{array}
\usepackage{dcolumn}
\usepackage{algorithm}
\usepackage{algpseudocode}

\usepackage{tikz}
\usetikzlibrary{shapes.geometric,arrows.meta,positioning,fit,backgrounds,calc}

\usepackage{xspace}
\usepackage{enumitem}
\usepackage{wrapfig}
\usepackage{url}
\usepackage{needspace}

\newcommand{\harness}{{\scshape HarnessDev}\xspace}
\newcommand{\hC}{{\scshape Creation}\xspace}
\newcommand{\hE}{{\scshape Evolution}\xspace}
\newcommand{\selfE}{{\scshape Self-Eval}\xspace}
\newcommand{\unifE}{{\scshape Unified-Eval}\xspace}

\definecolor{hexec}{HTML}{B0413E}
\definecolor{htool}{HTML}{3B8A4C}
\definecolor{hctx}{HTML}{3B6FB0}
\definecolor{hstate}{HTML}{B9A04F}
\definecolor{hlife}{HTML}{E07020}
\definecolor{heval}{HTML}{6F5BA8}
\definecolor{hmodel}{HTML}{222222}
\definecolor{lightgray}{HTML}{F2F2F2}
\definecolor{medgray}{HTML}{D9D9D9}
\definecolor{seedaccent}{HTML}{2E5AA8}
\definecolor{specbg}{HTML}{FAFAF8}
\newtcolorbox{specbox}{
  enhanced, breakable,
  colback=specbg,
  colframe=medgray,
  boxrule=0.5pt, arc=1.5mm,
  borderline west={2pt}{0pt}{seedaccent},
  left=3mm, right=2.5mm, top=0.5mm, bottom=0.5mm,
  fontupper=\small,
  before skip=8pt, after skip=8pt
}
\definecolor{accent}{HTML}{1F77B4}
\definecolor{tabhdr}{HTML}{2E2E2E}
\definecolor{tabrow}{HTML}{F4F6F9}

\title{\textsc{HarnessDev}: Can LLMs Create and Evolve Their Own Agent Harness?}

\affiliation[1]{ByteDance Seed}
\affiliation[2]{Singapore University of Technology and Design}
\affiliation[3]{\mbox{Georgia Institute of Technology}}
\affiliation[4]{M-A-P}
\affiliation[5]{TokenWave.AI}

\contribution{Full author list in Contributions}

\abstract{
As agents move from research prototypes to deployed tools, their
capability increasingly depends on model-external execution
infrastructure, commonly termed the agent \emph{harness}. Changing this
harness while holding model weights fixed can substantially alter task
performance. Current agent evaluations typically report downstream
performance under a chosen harness, leaving a model's ability to develop
the harness itself comparatively underexplored. We introduce \harness, a
benchmark that shifts the unit of evaluation from task outputs to
runnable infrastructure. \harness covers two stages. In
\emph{Creation}, the agent starts from a minimal seed and a small number
of cases, then builds a complete execution system. In \emph{Evolution},
it starts from its own created harness and iteratively revises it using
downstream execution feedback, with the goal of improving benchmark
performance. We then evaluate each constructed harness on \emph{capability}---task
success on held-out benchmarks, and \emph{efficiency}---execution-token
cost. The reported Creation results cover six creator LLMs, four domains,
and five downstream benchmarks totaling 2{,}207 unique downstream
instances, with hidden evaluation tasks withheld from development.
We find that generated harnesses remain substantially behind mature human-engineered
references on code and on search and research, while matching or exceeding
the selected references on writing and machine-learning experimentation,
with large variation in execution cost.
Evolution produces some performance gains, but they are unstable and
transfer only partially to held-out tasks. Experiments with a fixed
runtime model further show that the gains depend strongly on the model
executing the harness, indicating limited transfer across models.
}

\date{\today}

\checkdata[Project Page]{\url{https://self-developing-agents.github.io/}}

\begin{document}
\maketitle

\section{Introduction}
\label{sec:intro}

\begin{figure}[t]
\centering
\includegraphics[width=\linewidth]{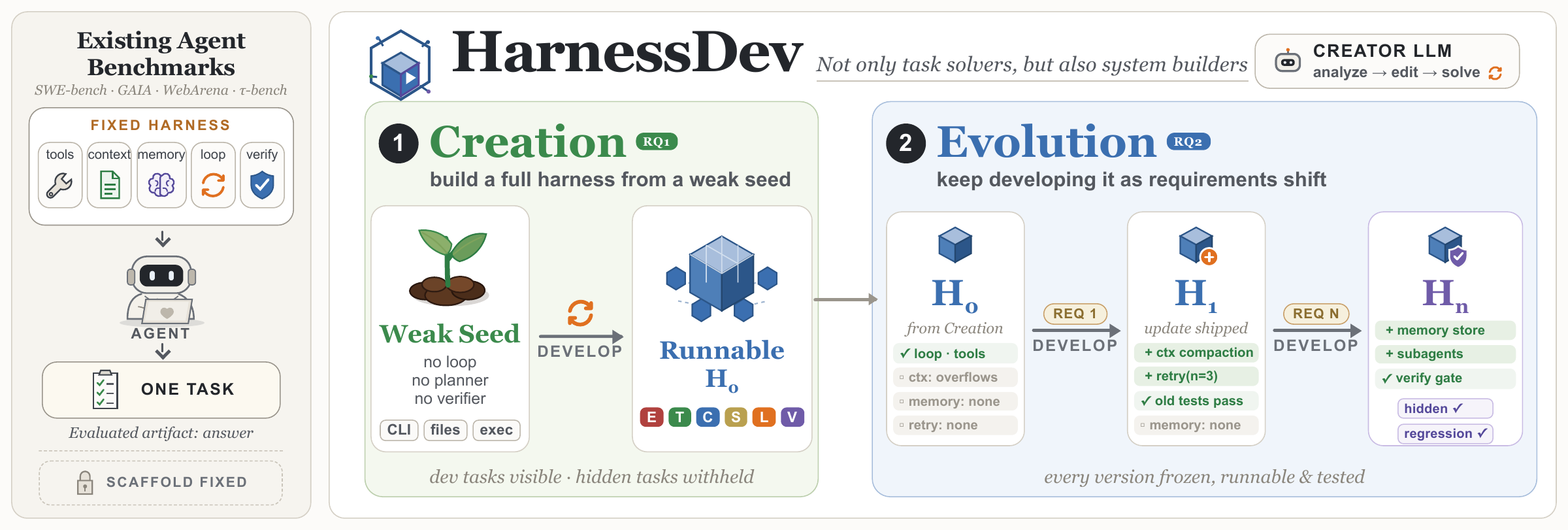}
\caption{\textbf{\textsc{HarnessDev} covers two stages of harness
development.} In \emph{Creation}, a creator builds a complete harness
from a weak but runnable seed and a small number of development cases.
In \emph{Evolution}, it continues to improve its own persistent harness
using downstream execution feedback. Both stages evaluate runnable
infrastructure that persists across tasks rather than a one-off task
output.}
\label{fig:teaser}
\end{figure}

As agents move from research prototypes to deployed tools such as coding
assistants~\citep{anthropic2024claude_code,openai2025codex}, data-analysis
copilots, browser workers~\citep{browseruse2026}, and research pipelines,
their capability increasingly depends on software outside the model's
weights. This surrounding execution infrastructure, commonly termed the
agent \emph{harness}~\citep{pan2026natural_language_harnesses,
ning2026code_agent_harness}, manages the execution loop, tool use, context,
failure recovery, and result verification that turn model outputs into
actions~\citep{anthropic2025context}. Its impact is substantial: with
identical weights, GPT-5 solves 35.2\% of Terminal-Bench~2.1 inside
Terminus~2 but 49.6\% inside Codex~CLI~\citep{tbench2026}. As agents
specialize to more domains, the demand for purpose-built harnesses will
continue to grow. Because these systems require continuous development
rather than one-time implementation, a practical question is whether LLMs
can assist harness engineers---or even take over such a role---in building
and continually improving the harness.

Despite this practical need, most agent evaluations select a harness for
a given comparison and report model performance on downstream
tasks~\citep{ICLR2024_edac78c3,ICLR2024_25ae35b5,ICLR2024_4410c071,
yao2024taubenchbenchmarktoolagentuserinteraction,ICLR2024_e9df36b2}.
This setup supports controlled task-level comparison, but treats the
harness as part of the experimental configuration rather than as an
artifact to be developed. Recent work has begun to study harness
representations, automated agent design, and agents that build or improve
agent systems~\citep{pan2026natural_language_harnesses,
ning2026code_agent_harness,ICLR2025_36b7acf6,ICLR2025_5492ecbc,
lu2026meta,zhang2026self}. However, it remains underexplored whether models
can both create and continually improve runnable, persistent harnesses.
Answering this question requires separating the model that develops the
harness from the model that executes downstream tasks, recording the
development environment, and measuring downstream performance, transfer
across executors, distance from human-engineered systems, regression, and
cost.

This evaluation gap is particularly consequential because harness
engineering is fundamentally different from ordinary code editing.
When a model modifies a standalone program, the target behavior is
externally specified and success is locally verifiable. When a model
modifies its own harness, it is editing the execution substrate
through which it acts: the change alters how the model itself
observes, plans, and recovers in all future tasks. Effective harness
improvement therefore demands that the model recognize its own
behavioral limitations from execution
traces~\citep{shinn2023reflexionlanguageagentsverbal}, diagnose
structural bottlenecks in the system it runs inside, and commit
targeted changes that accumulate into lasting, reusable capability
gains rather than one-off
fixes~\citep{NEURIPS2024_5a7c9475,wang2025openhandsopenplatformai}.
As frontier models grow capable enough to edit multi-file codebases
and close real pull requests, this ability is already latent; what is
missing is a benchmark that measures it.

We introduce \textbf{\textsc{HarnessDev}} (Figure~\ref{fig:teaser}), a
benchmark that fills this gap by shifting the unit of evaluation from
task outputs to runnable infrastructure: measuring a model's ability
to construct and maintain execution systems that are durable,
inspectable, and reusable. The name reflects the software-engineering
sense of \emph{develop}: developing a harness includes both building
it from scratch (i.e., Creation) and improving it through continued iteration and
maintenance (i.e., Evolution).
The benchmark covers two stages of harness development. In
\emph{Creation}, a creator LLM starts from a deliberately weak but
runnable seed and builds a complete harness for a new task family. In
\emph{Evolution}, it starts from an existing harness and continues to
develop it toward better downstream task performance. Together, the two
stages evaluate whether models can complete harness development tasks
and continuously improve the resulting system.

Evaluating a generated harness is harder than evaluating a generated
answer. A harness can overfit to the model that wrote it, memorize
development examples, improve one capability while silently regressing
another, or improve the feedback-set score through benchmark-specific
changes that do not transfer to new tasks. We therefore evaluate along
two axes.
\emph{Capability} measures whether the harness works: we run it on
held-out downstream tasks and report task-level success rates.
\emph{Efficiency} measures how many executor-model tokens the frozen
harness consumes when deployed to solve downstream tasks.

Our findings follow the two stages of harness development: 

\paragraph{Harness Creation.}
Current models can construct runnable harnesses from a weak seed, but the
gap from mature human-engineered systems varies substantially by harness
type. When each harness runs with the model that built it, model-built
harnesses match the reference on short-form writing and exceed it on
machine-learning experimentation. The gap is largest for search and
research harnesses, which require long-horizon information seeking, and
remains substantial for code harnesses, which must coordinate repository
inspection, editing, and verification over many turns. Harnesses
built by different creator models also differ substantially not only in
downstream task performance, but also in the number of executor tokens they
consume. Higher execution cost does not reliably produce better results,
so harness quality must be assessed through both capability and efficiency.

\paragraph{Harness Evolution.}
Current models can use downstream execution feedback to improve their
own harnesses, but reliable evolution remains difficult. Performance
often rises and falls across successive revisions, and gains observed
during development become smaller and less consistent on unseen tasks.
The outcome also depends strongly on the model that runs the harness:
changing this runtime model alters both the starting performance and
whether subsequent revisions help. These results show that models can
make useful local improvements, while robust evolution across unseen
tasks and runtime models remains an open challenge.

\section{Background}
\label{sec:background}
Most agent benchmarks begin after the problem has already been made executable: the
task is specified, the reward or judge is defined, and the execution scaffold is
fixed. This setting is necessary for controlled comparison, but it hides the work
that dominates real deployment. In industry that work is spread across several
roles---solutions architects, applied and platform engineers---but its most
visible recent crystallization is the \emph{forward-deployed engineer} (FDE), a
title popularized by Palantir and since adopted by frontier-model
companies~\citep{palantir2020fdse,orosz2025fde}. An FDE is embedded at the
customer site after a system is adopted and turns a general-purpose model into
something that runs against that customer's data formats, workflows, and
compliance constraints---for example, rewriting an ingestion path because logs
may only be retained for a fixed period, or localizing a failure in a
cross-jurisdiction contract pipeline. The role's success criterion is not a
demo or a benchmark score but whether the deployed system is genuinely used,
keeps working, and improves; its failures are folded back into the product as
fixes and feature requests~\citep{orosz2025fde}. The rapid growth of FDE hiring
across frontier-model and data-platform companies~\citep{thenewstack2025fde}
reflects a simple fact: a capable model is not yet a working
system~\citep{nanda2025state}, and today the gap is closed by human
engineers.

Viewed from the model's side, FDE work supplies three pieces of structure that
benchmark designers normally presuppose. First, the target is vague: an informal
business intent must be translated into concrete objectives, constraints, and
success criteria. Second, the feedback signal is absent or unreliable: tests,
judges, traces, or other self-evaluation must be constructed before anyone can
tell whether the system is improving---``compliant'' only becomes checkable once
someone encodes what compliance means here. Third, the execution system does not
exist in a usable form: the tools, context management, state, lifecycle logic,
and verification interface through which future tasks will run must be built,
adapted, and then \emph{maintained}---an FDE stays with the system as
requirements shift, rather than delivering once and leaving.

This paper focuses on the third layer. In a typical controlled agent
evaluation, researchers select an agent configuration and report task
completion under that configuration. The harness is therefore usually part of
the evaluation setup rather than the object being developed. \textsc{HarnessDev}
instead asks whether language models can create this execution scaffold from a
weak starting point and then improve it using feedback while preserving
constraint compliance and held-out performance. Whereas \textsc{Aspire} studies
how broad deployment needs become capability growth and S$^3$Gym studies whether
interaction experience can be judged and reused, \textsc{HarnessDev} isolates
how models build and maintain the systems that carry them.

\FloatBarrier
\Needspace{8\baselineskip}
\section{Benchmark}
\label{sec:data}

\subsection{Overview}
\label{sec:taskfamilies}

\harness evaluates the execution system that a model develops, rather
than the answer it produces for a single task. The submitted artifact is
a runnable harness that is frozen and then reused across downstream
tasks. A creator LLM $L_C$ works inside a development environment $D$
to produce a runnable harness $H$. The
development signal differs by setting and is defined in
Table~\ref{tab:devsettings}. After development, $H$ is frozen. An
executor LLM $L_E$ then runs inside it on a downstream task $x$, and
evaluator $J$ scores the resulting output $y$:
\begin{equation}
(L_C,D)\rightarrow H,
\qquad
(H,L_E,x)\rightarrow y\xrightarrow{J}\mathrm{score}.
\end{equation}
Thus, $D$ is used to build $H$, whereas $L_E$ is used only after $H$ is
frozen. In implementation terms, $H$ contains the execution loop, tools,
context management, persistent state, lifecycle control, and verification;
we describe these components in words rather than assigning each another
symbol.

The benchmark studies two stages of harness development.
\paragraph{RQ1---Creation.}
Can a model build an effective harness from a weak but runnable seed?
The creator must turn a task specification and a few development cases
into infrastructure that generalizes to unseen tasks.

\paragraph{RQ2---Evolution.}
Can a model improve an existing harness while preserving behavior that
already works? The creator evolves its own Creation harness from
downstream execution feedback. We additionally analyze the resulting
artifacts and trajectories, including edit statistics, feedback response,
held-out generalization, and transfer across executors.

\subsection{Development settings}
\label{sec:settings}

All settings provide a mutable development workspace, but they differ
in the starting harness and the signal available to the creator.
Table~\ref{tab:devsettings} gives the central distinction.

\begin{table}[h!]
\centering
\small
\setlength{\tabcolsep}{4pt}
\renewcommand{\arraystretch}{1.08}
\begin{tabular}{@{}>{\raggedright\arraybackslash}p{0.10\textwidth}
                    >{\raggedright\arraybackslash}p{0.16\textwidth}
                    >{\raggedright\arraybackslash}p{0.34\textwidth}
                    >{\raggedright\arraybackslash}p{0.32\textwidth}@{}}
\toprule
\textbf{Setting} & \textbf{Starts from} & \textbf{Development signal} & \textbf{Output} \\
\midrule
Creation & Weak seed $H_{\mathrm{seed}}$ & Specification and 1--3 development cases
& Final harness $H$ \\
Evolution & Creator's RQ1 $H_0$ & Results from a designated feedback set
& Frozen paired candidates and a creator-declared final harness \\
\bottomrule
\end{tabular}
\caption{\textbf{Harness-development settings.} Creation builds a new
harness; Evolution improves that harness from downstream execution feedback.}
\label{tab:devsettings}
\end{table}

\paragraph{Weak seed $H_{\mathrm{seed}}$.}
Creation should measure whether a model can design an execution system,
not whether it can reproduce benchmark boilerplate. Every creator
therefore receives the same $H_{\mathrm{seed}}$: a runnable
compatibility layer, not a task-solving agent. It parses task and model
configuration, exposes permitted low-level tools, and writes the
required results, trajectories, logs, and task artifacts. Its tools are
passive and act only when the harness calls them.

The seed has no agent loop, task decomposition, tool policy, context
management, persistent task state, verifier, retry or recovery logic,
or stopping rule. It may issue one connectivity probe, but it does not
attempt the task. Unmodified, it produces an empty or partial artifact
and scores zero on every downstream benchmark. Any nonzero Creation
score must therefore come from execution logic added by the creator.
This boundary matters because real scaffolds combine many control
primitives, and their composition affects task
performance~\citep{pan2026natural_language_harnesses,
ning2026code_agent_harness,
rombaut2026insidescaffoldsourcecodetaxonomy}.

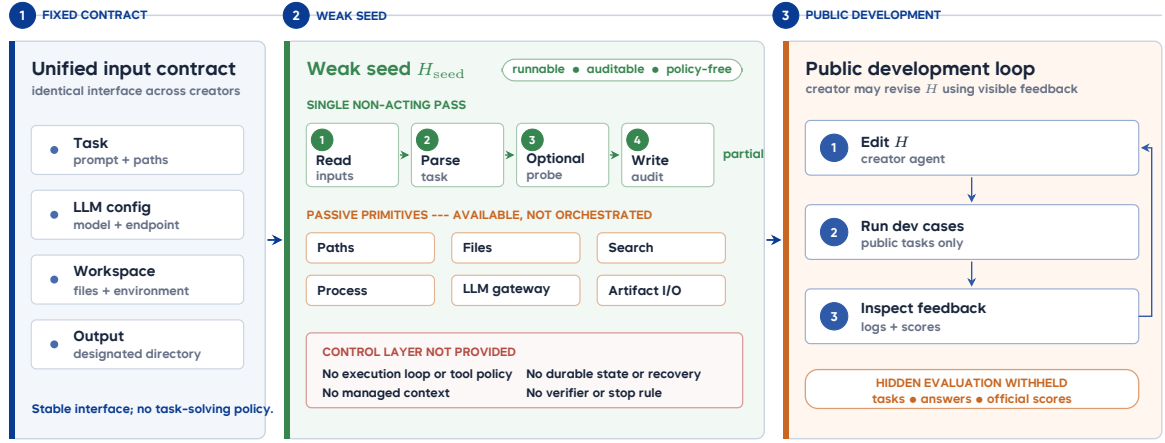
\begin{figure}[t]
\centering
\resizebox{\linewidth}{!}{
\definecolor{cNavy}{HTML}{083B91}
\definecolor{cBlue}{HTML}{2E5AA8}
\definecolor{cBluePale}{HTML}{F3F7FC}
\definecolor{cGreen}{HTML}{378650}
\definecolor{cGreenPale}{HTML}{F2F8F4}
\definecolor{cOrange}{HTML}{C96722}
\definecolor{cOrangePale}{HTML}{FFF7EF}
\definecolor{cRed}{HTML}{B64B43}
\definecolor{cInk}{HTML}{17243A}
\definecolor{cMuted}{HTML}{667085}
\definecolor{cRule}{HTML}{CBD5E1}

\begin{tikzpicture}[x=1cm,y=1cm,every node/.style={inner sep=0pt}]
  \tikzset{
    stage/.style={font=\fontsize{7.4}{8.5}\sffamily\bfseries,text=cInk},
    overline/.style={font=\fontsize{4.7}{5.6}\sffamily\bfseries,text=cMuted},
    body/.style={font=\fontsize{5.7}{6.7}\sffamily,text=cInk,align=left},
    tiny/.style={font=\fontsize{4.8}{5.8}\sffamily,text=cMuted,align=left},
    panel/.style={draw=cRule,rounded corners=3pt,line width=.55pt,fill=white},
    card/.style={draw=cRule,rounded corners=2pt,line width=.45pt,fill=white},
    arrow/.style={-{Stealth[length=4.2pt,width=4.2pt]},line width=.7pt,draw=cNavy},
    thinarr/.style={-{Stealth[length=3.5pt,width=3.5pt]},line width=.55pt,draw=cBlue},
  }

  \draw[cRule,line width=.6pt] (0,.18) -- (16,.18);
  \foreach \x/\n/\lbl in {0.18/1/{FIXED CONTRACT},3.98/2/{WEAK SEED},10.78/3/{PUBLIC DEVELOPMENT}}{
    \node[draw=cNavy,fill=cNavy,text=white,circle,minimum size=.36cm,
      font=\fontsize{5.2}{5.2}\sffamily\bfseries] at (\x,.18) {\n};
    \node[overline,anchor=west,text=cNavy] at (\x+.27,.18) {\lbl};
  }

  \draw[panel,fill=cBluePale]   (0,-.18) rectangle (3.55,-5.70);
  \draw[panel,fill=cGreenPale]  (3.82,-.18) rectangle (10.48,-5.70);
  \draw[panel,fill=cOrangePale] (10.75,-.18) rectangle (16,-5.70);

  \fill[cNavy] (0,-.18) rectangle (.07,-5.70);
  \node[stage,anchor=west] at (.30,-.58) {Unified input contract};
  \node[tiny,anchor=west] at (.30,-.86) {identical interface across creators};

  \foreach \y/\ttl/\sub in {
    -1.35/{Task}/{prompt + paths},
    -2.25/{LLM config}/{model + endpoint},
    -3.15/{Workspace}/{files + environment},
    -4.05/{Output}/{designated directory}}{
    \draw[card] (.30,\y) rectangle (3.25,\y-.68);
    \fill[cNavy!75] (.62,\y-.34) circle (.055);
    \node[body,font=\fontsize{5.7}{6.5}\sffamily\bfseries,anchor=west]
      at (.88,\y-.25) {\ttl};
    \node[tiny,anchor=west] at (.88,\y-.49) {\sub};
  }
  \node[tiny,anchor=west,text=cNavy] at (.30,-5.30) {Stable interface; no task-solving policy.};

  \fill[cGreen] (3.82,-.18) rectangle (3.89,-5.70);
  \node[stage,anchor=west,text=cGreen] at (4.12,-.58) {Weak seed $H_{\mathrm{seed}}$};
  \node[draw=cGreen!55,fill=white,rounded corners=5pt,line width=.45pt,
    font=\fontsize{4.7}{5.4}\sffamily\bfseries,text=cGreen,inner xsep=4pt,inner ysep=2pt,
    anchor=east] at (10.18,-.58) {runnable \;\textbullet\; auditable \;\textbullet\; policy-free};

  \node[overline,anchor=west,text=cGreen] at (4.12,-1.06) {SINGLE NON-ACTING PASS};
  \foreach \x/\num/\a/\b in {
    4.12/1/{Read}/{inputs},
    5.58/2/{Parse}/{task},
    7.04/3/{Optional}/{probe},
    8.50/4/{Write}/{audit}}{
    \draw[card,draw=cGreen!45] (\x,-1.32) rectangle (\x+1.28,-2.20);
    \node[draw=cGreen,fill=cGreen,text=white,circle,minimum size=.30cm,
      font=\fontsize{4.5}{4.5}\sffamily\bfseries] at (\x+.22,-1.55) {\num};
    \node[body,font=\fontsize{5.4}{6.2}\sffamily\bfseries,anchor=west]
      at (\x+.13,-1.83) {\a};
    \node[tiny,anchor=west] at (\x+.13,-2.06) {\b};
  }
  \draw[thinarr,draw=cGreen] (5.42,-1.76) -- (5.55,-1.76);
  \draw[thinarr,draw=cGreen] (6.88,-1.76) -- (7.01,-1.76);
  \draw[thinarr,draw=cGreen] (8.34,-1.76) -- (8.47,-1.76);
  \node[tiny,text=cGreen,anchor=west] at (9.90,-1.76) {partial};

  \node[overline,anchor=west,text=cOrange] at (4.12,-2.60) {PASSIVE PRIMITIVES --- AVAILABLE, NOT ORCHESTRATED};
  \foreach \x/\y/\txt in {
    4.12/-2.84/{Paths},6.14/-2.84/{Files},8.16/-2.84/{Search},
    4.12/-3.43/{Process},6.14/-3.43/{LLM gateway},8.16/-3.43/{Artifact I/O}}{
    \draw[card,draw=cOrange!50,fill=white] (\x,\y) rectangle (\x+1.78,\y-.42);
    \node[body,font=\fontsize{5.1}{5.9}\sffamily,anchor=west] at (\x+.15,\y-.21) {\txt};
  }

  \draw[draw=cRed!55,fill=cRed!4,rounded corners=2pt,line width=.5pt]
    (4.12,-4.23) rectangle (10.18,-5.27);
  \node[overline,anchor=west,text=cRed] at (4.34,-4.50) {CONTROL LAYER NOT PROVIDED};
  \node[tiny,anchor=west,text=cInk] at (4.34,-4.82)
    {No execution loop or tool policy};
  \node[tiny,anchor=west,text=cInk] at (7.17,-4.82)
    {No durable state or recovery};
  \node[tiny,anchor=west,text=cInk] at (4.34,-5.08)
    {No managed context};
  \node[tiny,anchor=west,text=cInk] at (7.17,-5.08)
    {No verifier or stop rule};

  \fill[cOrange] (10.75,-.18) rectangle (10.82,-5.70);
  \node[stage,anchor=west,text=cInk] at (11.05,-.58) {Public development loop};
  \node[tiny,anchor=west] at (11.05,-.86) {creator may revise $H$ using visible feedback};

  \foreach \y/\num/\ttl/\sub in {
    -1.28/1/{Edit $H$}/{creator agent},
    -2.45/2/{Run dev cases}/{public tasks only},
    -3.62/3/{Inspect feedback}/{logs + scores}}{
    \draw[card,draw=cBlue!50] (11.05,\y) rectangle (15.68,\y-.76);
    \node[draw=cBlue,fill=cBlue,text=white,circle,minimum size=.38cm,
      font=\fontsize{5.0}{5.0}\sffamily\bfseries] at (11.45,\y-.38) {\num};
    \node[body,font=\fontsize{5.7}{6.5}\sffamily\bfseries,anchor=west]
      at (11.82,\y-.29) {\ttl};
    \node[tiny,anchor=west] at (11.82,\y-.55) {\sub};
  }
  \draw[thinarr] (13.36,-2.07) -- (13.36,-2.42);
  \draw[thinarr] (13.36,-3.24) -- (13.36,-3.59);
  \draw[thinarr] (15.68,-4.00) -- (15.86,-4.00) -- (15.86,-1.66) -- (15.71,-1.66);

  \draw[draw=cOrange!65,fill=white,rounded corners=5pt,line width=.5pt]
    (11.05,-4.74) rectangle (15.68,-5.28);
  \node[overline,text=cOrange] at (13.365,-4.92) {HIDDEN EVALUATION WITHHELD};
  \node[tiny,text=cOrange] at (13.365,-5.15) {tasks \textbullet\ answers \textbullet\ official scores};

  \draw[arrow] (3.58,-2.94) -- (3.79,-2.94);
  \draw[arrow] (10.51,-2.94) -- (10.72,-2.94);
\end{tikzpicture}}
\caption{\textbf{The weak Seed Harness and its development
environment.} The seed fixes only the input and audit envelopes and
exposes passive, unorchestrated primitives; unmodified, it performs no
task work. Development feedback is public; hidden tasks, answers, and
official scores remain inaccessible.}
\label{fig:weak-seed-env}
\end{figure}

\begin{figure}[b!]
\centering
\resizebox{\linewidth}{!}{
\definecolor{cNavy}{HTML}{083B91}
\definecolor{cBlue}{HTML}{2E5AA8}
\definecolor{cBluePale}{HTML}{F3F7FC}
\definecolor{cInk}{HTML}{17243A}
\definecolor{cMuted}{HTML}{667085}
\definecolor{cRule}{HTML}{CBD5E1}

\begin{tikzpicture}[x=1cm,y=1cm,every node/.style={inner sep=0pt}]
  \tikzset{
    title/.style={font=\fontsize{7.6}{8.7}\sffamily\bfseries,text=cInk},
    label/.style={font=\fontsize{5.4}{6.3}\sffamily\bfseries,text=cInk},
    body/.style={font=\fontsize{5.2}{6.1}\sffamily,text=cInk,align=left},
    tiny/.style={font=\fontsize{4.7}{5.6}\sffamily,text=cMuted,align=left},
    shell/.style={draw=cNavy,rounded corners=3pt,line width=.8pt,fill=white},
    card/.style={draw=cRule,rounded corners=2pt,line width=.45pt,fill=white},
    arrow/.style={-{Stealth[length=4.2pt,width=4.2pt]},line width=.75pt,draw=cNavy},
  }

  \draw[draw=hexec!70,fill=hexec!4,rounded corners=3pt,line width=.65pt]
    (0,-1.64) rectangle (2.28,-4.16);
  \node[draw=hexec,fill=white,circle,minimum size=.75cm,line width=.6pt,
    font=\fontsize{8.0}{8.0}\sffamily\bfseries,text=hexec] at (1.14,-2.18) {C};
  \node[title,text=hexec] at (1.14,-2.82) {Creator LLM};
  \node[body,align=center,text width=1.92cm] at (1.14,-3.25)
    {implements the harness control layer};
  \node[tiny,align=center,text width=1.92cm] at (1.14,-3.82)
    {not the downstream task answer};
  \draw[arrow,draw=hexec] (2.30,-2.90) -- (2.78,-2.90);

  \draw[shell,fill=cBluePale] (2.82,-.18) rectangle (11.82,-5.62);
  \node[title,text=cNavy,anchor=west] at (3.12,-.52)
    {Created runnable harness $H$};
  \node[tiny,anchor=east,text=cNavy] at (11.52,-.52)
    {one integrated executable system};

  \node[tiny,anchor=west,text=cMuted] at (3.12,-.91) {CONTROL PLANE IMPLEMENTED BY THE CREATOR};
  \foreach \x/\cc/\letter/\name in {
    3.12/hexec/E/{Loop},
    4.49/htool/T/{Tools},
    5.86/hctx/C/{Context},
    7.23/hstate/S/{State},
    8.60/hlife/L/{Lifecycle},
    9.97/heval/V/{Verify}}{
    \draw[draw=\cc!75!black,fill=\cc!7,rounded corners=2pt,line width=.55pt]
      (\x,-1.14) rectangle (\x+1.20,-2.18);
    \node[font=\fontsize{8.5}{8.5}\sffamily\bfseries,text=\cc]
      at (\x+.60,-1.50) {\letter};
    \node[body,font=\fontsize{4.8}{5.5}\sffamily\bfseries,text=\cc]
      at (\x+.60,-1.91) {\name};
  }

  \foreach \x in {3.72,5.09,6.46,7.83,9.20,10.57}{
    \draw[-{Stealth[length=3.2pt,width=3.2pt]},draw=cRule!80!cNavy,line width=.45pt]
      (\x,-2.19) -- (\x,-2.48);
  }
  \draw[draw=cNavy!70,fill=cNavy,rounded corners=2pt,line width=.55pt]
    (3.12,-2.50) rectangle (11.52,-3.30);
  \node[label,text=white] at (7.32,-2.76) {Harness execution core};
  \node[tiny,text=white!88] at (7.32,-3.08)
    {the six mechanisms jointly govern downstream task execution};

  \draw[card,draw=cBlue!55] (3.12,-3.62) rectangle (7.22,-5.22);
  \node[label,text=cBlue,anchor=west] at (3.36,-3.92) {Unified audit contract};
  \node[body,anchor=west] at (3.36,-4.32) {\texttt{result.json}\quad \texttt{trajectory.jsonl}};
  \node[body,anchor=west] at (3.36,-4.70) {\texttt{response.md}\quad runtime logs};
  \node[tiny,font=\fontsize{4.3}{5.1}\sffamily,text=cMuted,align=left,anchor=west] at (3.36,-5.02) {shared status, metrics, actions, and observations};

  \draw[card,draw=cBlue!55] (7.48,-3.62) rectangle (11.52,-5.22);
  \node[label,text=cBlue,anchor=west] at (7.72,-3.92) {Domain finalizer};
  \node[body,anchor=west] at (7.72,-4.32) {maps the run into the domain's};
  \node[body,anchor=west] at (7.72,-4.64) {authoritative scorer-readable artifact};
  \node[tiny,font=\fontsize{4.3}{5.1}\sffamily,text=cMuted,align=left,anchor=west] at (7.72,-5.02) {the shared contract does not force one file format};

  \draw[arrow] (11.54,-4.42) -- (12.12,-4.42);
  \draw[cNavy,line width=.75pt] (12.12,-1.30) -- (12.12,-4.99);

  \node[title,text=cNavy,anchor=west] at (12.42,-.52) {Scorer-readable artifacts};
  \node[tiny,anchor=west] at (12.42,-.83) {authoritative output varies by domain};

  \foreach \y/\name/\desc in {
    -1.05/{Code}/{repository state + patch},
    -2.12/{Data / MLE}/{submission + metrics},
    -3.19/{Writing}/{final user-facing prose},
    -4.26/{Search}/{answer + cited evidence}}{
    \draw[cNavy,line width=.65pt] (12.12,\y-.42) -- (12.38,\y-.42);
    \draw[card,draw=cNavy!45,fill=white] (12.40,\y) rectangle (16,\y-.84);
    \fill[cNavy] (12.40,\y) rectangle (12.48,\y-.84);
    \node[label,text=cNavy,anchor=west] at (12.70,\y-.27) {\name};
    \node[tiny,anchor=west,text=cInk] at (12.70,\y-.59) {\desc};
  }

  \node[tiny,text=cNavy,anchor=west] at (12.42,-5.43)
    {one output bus; no cross-domain file-format assumption};
\end{tikzpicture}}
\caption{\textbf{From control layer to scorable artifacts.} The
creator must implement the six control modules ($E/T/C/S/L/V$, colored
as in Figure~\ref{fig:teaser}). A finished harness reports through
unified auditable outputs and delivers a domain-specific,
scorer-readable final artifact.}
\label{fig:creator-contract}
\end{figure}
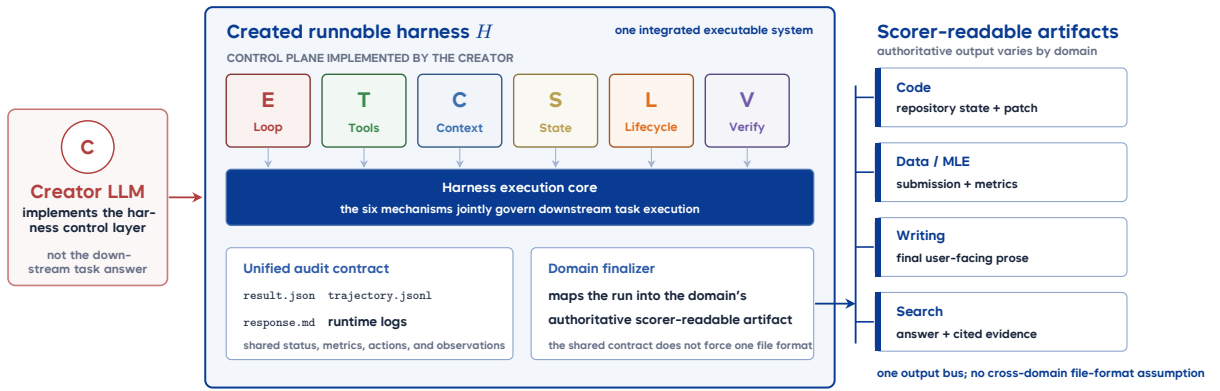

This design avoids two extremes. An empty repository would mix harness
design with command-line and file-format setup; a mature agent would
give away the planning and verification structure being tested.
$H_{\mathrm{seed}}$ removes the setup burden without providing a
solution policy. Figure~\ref{fig:weak-seed-env} shows the seed and its
development environment; Figure~\ref{fig:creator-contract} shows the
control layer the creator must implement and the scorer-readable
artifacts a finished harness delivers.
Appendix~\ref{sec:appseed} gives its implementation skeleton.

\paragraph{Creation (RQ1).}
Along with $H_{\mathrm{seed}}$, the creator receives a task-family
specification, tool and permission constraints, a short design tutorial,
and one to three development cases. It may revise the harness using
feedback from those cases, but it never sees the human implementation or
the hidden evaluation set. The resulting harness $H$ is frozen before
evaluation.

\paragraph{Evolution (RQ2).}
The creator starts from its own frozen RQ1 code harness $H_0$. During
development, it receives results from a fixed 100-task SWE-Pro feedback
set and all 89 Terminal-Bench tasks. The 100 SWE-Pro tasks are a subset
of the 731-instance public split used in Creation, and the 630-instance
held-out split of Section~\ref{sec:exp-rq2} is drawn from the same split.
In the reported protocol,
the controller first evaluates $H_0$ on both benchmarks. Each official
post-$H_0$ candidate is then frozen and submitted as a pair: one complete
100-task SWE-Pro evaluation and one complete 89-task Terminal-Bench
evaluation of the same commit. A candidate enters the official trajectory
only after both legs settle. Same-commit infrastructure repairs are merged;
probes, partial legs, stopped runs, and invalid instances are excluded.

The controller provides a budget of ten post-$H_0$ full-evaluation pairs. Between two
charged pairs, the creator may use at most two fixed-subset probes, each
covering the same first five tasks from both benchmarks. Probe results are
diagnostic and never become official scores. The creator terminates by
declaring a non-$H_0$ commit that has a complete official pair.

Both benchmarks shown during Evolution are feedback-bearing development
sets, so in-trajectory scores measure online adaptation and version
selection. Generalization is measured separately after freezing: every
official version is additionally evaluated on 630 SWE-Pro instances
disjoint from the feedback set, and these scores are never shown to the
creator.
Throughout this paper, \emph{held-out} means withheld from the creator's
development loop; Section~\ref{sec:exp-rq2} gives the full setting.

\subsection{Domains and downstream benchmarks}
\label{sec:categories}
\label{sec:corpus}

Creation covers four domains and five downstream benchmarks
(Table~\ref{tab:domains}); Evolution currently focuses on code
harnesses. Together, the suites contain 2{,}207 unique downstream
instances. The Evolution feedback tasks come from the same benchmark suites
and are therefore not counted again.

\begin{table}[h!]
\centering
\small
\setlength{\tabcolsep}{5pt}
\renewcommand{\arraystretch}{1.06}
\begin{tabular}{@{}>{\raggedright\arraybackslash}p{0.21\textwidth}
                    >{\raggedright\arraybackslash}p{0.35\textwidth}
                    >{\centering\arraybackslash}p{0.09\textwidth}
                    >{\raggedright\arraybackslash}p{0.25\textwidth}@{}}
\toprule
\textbf{Domain} & \textbf{Benchmark} & \textbf{Tasks} & \textbf{Primary metric} \\
\midrule
Code & SWE-bench Pro~\citep{deng2025swebenchpro}, public split
(abbreviated SWE-Pro below) & 731 & Task success \\
Code & Terminal-Bench~2.1~\citep{tbench2026} & 89 & Task success \\
Data analysis & MLE-bench~\citep{chan2025mlebench}
& 75 & Medal score \\
Writing & EQ-Bench3~\citep{eqbench2025}
& 46 & Rubric score \\
Research & BrowseComp~\citep{wei2025browsecomp}
& 1{,}266 & Accuracy \\
\bottomrule
\end{tabular}
\caption{\textbf{Downstream evaluation coverage for Creation.} Creation
uses all 731 instances of the SWE-bench Pro public split. Evolution draws
its 100-task SWE-Pro feedback set and its 630-instance held-out split
from that same public split, and uses all 89 Terminal-Bench tasks as
feedback.}
\label{tab:domains}
\end{table}

Mature open-source systems define the capability surface in each
domain. Depending on availability, a system may serve as a
human-engineered reference or provide a development environment. These
roles are assigned separately; inclusion does not imply that a system
serves both. Appendix~\ref{sec:appcorpus}
lists the candidate systems, while the benchmark release fixes their
roles, versions, and licenses.

\FloatBarrier

\subsection{Evaluation protocol}
\label{sec:eval}

Every score is produced by a frozen harness in a standardized runtime.
The executor LLM $L_E$ and evaluator $J$ remain fixed within each
comparison, so score changes reflect changes to the harness.
Development and evaluation are also separated: hidden scores are not
returned in Creation, and Evolution exposes only its designated feedback
set during development.

\paragraph{Creation.}
We compare a created harness with the common seed and, where available,
a mature human-engineered harness. \selfE{} sets $L_E=L_C$ and measures
the complete creator--harness system. \unifE{} runs every generated
harness with the same fixed $L_E$, making harnesses directly comparable.

\paragraph{Evolution.}
Evolution candidates are evaluated on the designated feedback benchmarks
during development. Only complete two-benchmark pairs enter the official
trajectory, and the creator selects a final paired candidate. After all
trajectories end, every official version is additionally evaluated on a
disjoint held-out set that is never shown to the creator, so adaptation
to observed feedback and held-out generalization are reported
separately. Appendix~\ref{sec:appevalmetrics} details the evaluation
settings and model roles.

\paragraph{Constraint compliance.}
The creator-visible specification states what a submitted harness may
not do: hard-code instance-specific solutions, derive patches from task
identifiers, file-name allowlists, or known answers, consult hidden
tests, hidden answers, hidden patches, private scorer internals, or
official evaluation feedback, or replace the provided provider-neutral
runtime interface with its own LLM access path. Two properties make
these constraints checkable rather than advisory. First, the score path
is isolated from the harness: a harness's self-reported status is never
a scoring input, SWE-Pro credit comes only from the real repository diff
left in the task workdir, and Terminal-Bench credit only from final
environment state, so no harness can earn score by asserting success.
Second, every run retains its trajectory, result, and metric artifacts
alongside the frozen harness source, which supports a post-hoc audit of
the delivered code and of what that code actually executed. We audited
the delivered harness source and the recorded execution artifacts of
every run reported in this paper, and report the outcome as a null
result: no harness obtained score through a prohibited route, and no run
is excluded on these grounds.

\subsection{Metrics}
\label{sec:metrics}%

For each frozen harness, we report two quantities: downstream task
performance under the benchmark's native metric and the executor-model
tokens consumed during evaluation. Execution cost is reported as both
the total and the mean per task; tokens used by the creator to build or
modify the harness are excluded. Creation compares each harness with the
weak seed and available human-engineered references. Evolution reports
the performance change from its initial harness $H_0$ under the same
executor and scorer.

\section{Experiments}
\label{sec:exp}

We instantiate the benchmark defined in Section~\ref{sec:data} and
report results for Creation, Evolution, and cross-model behavior. The
behavioral analysis compares resource use, response to feedback, and
transfer across executors.

\subsection{Experimental setup}
\label{sec:exp-setup}

We evaluate six creator LLMs $L_C$:
Opus~4.8~\citep{anthropic2026opus}, GPT-5.5~\citep{openai2026gpt55},
Gemini~3.1~Pro~\citep{google2026gemini31},
DeepSeek~V4~Pro~\citep{deepseek2026v4},
Qwen~3.7~Max~\citep{alibaba2026qwen37}, and
Seed~2.0~Pro~\citep{bytedance2026seed20}. Models run through their
official APIs or OpenRouter~\citep{openrouter2026}. We use Claude
Code~2.1.177 as the development environment $D$, except that GPT-5.5
uses Codex~0.144.3. Appendix~\ref{sec:apptables} gives the model,
decoding, and development-environment configuration, the downstream
execution resources and time limits, and the human-reference sources.

We follow the development and evaluation protocols in
Sections~\ref{sec:settings}--\ref{sec:eval}. For RQ1, we independently
create and evaluate three harnesses for each creator--benchmark pair and
report avg@3. Under \selfE{}, the creator also serves as executor;
under \unifE{}, every harness uses Gemini~3.1~Pro, which isolates
executor compatibility. Human-reference results are verified public
system results rather than paired controls under one executor; their
sources are listed in Appendix~\ref{sec:humanrefs}.

\subsection{Harness Creation}
\label{sec:exp-rq1}

RQ1 asks whether a model can turn the runnable weak seed into an
effective harness for a task family. We compare the generated harnesses
with the zero-scoring seed and with verified human-engineered systems,
reporting both downstream performance and execution tokens.

\begin{figure}[tbp]
\centering
\includegraphics[width=\linewidth]{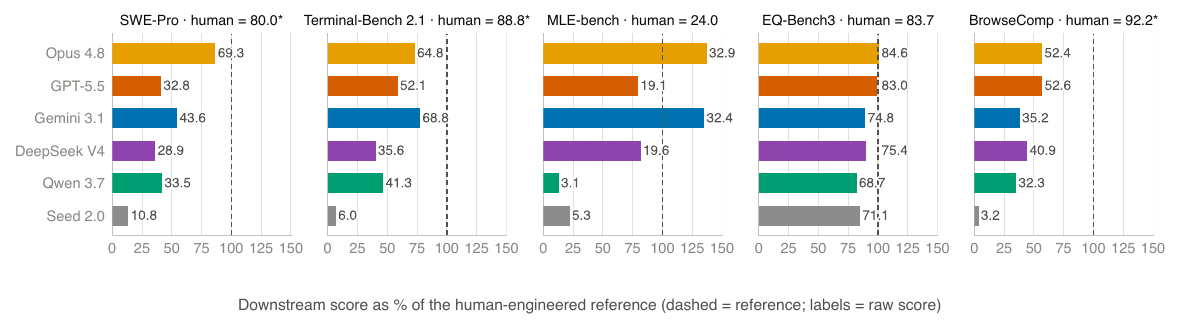}
\caption{\textbf{RQ1: distance to human-engineered references.}
Each creator's \selfE{} downstream score is normalized to the human-engineered reference for that benchmark; the dashed line is the reference and the label beside each bar is the raw score. Human references come from different harness--model combinations and are not paired controls under a common executor, so this figure only shows the distance to the selected mature systems. Exceeding $100\%$ means exceeding that external reference, not exceeding human ability.}
\label{fig:rq1gap}
\end{figure}

\paragraph{Overall findings.}
Creation quality varies substantially across task families
(Tables~\ref{tab:rq1self}--\ref{tab:rq1unified} and
Figure~\ref{fig:rq1gap}). Under \selfE{}, Opus~4.8 has the highest
overall score (67.8), but remains below the human-engineered reference
(86.2). Writing harnesses approach the reference, whereas Search shows
the largest gap and Code also remains behind. Opus~4.8 and
Gemini~3.1~Pro lead MLE-bench with medal rates of 32.9 and 32.4.
More broadly, $77.8\%$ of failed Data tasks are attributed to harness
defects, showing that the bottleneck is not only executor capability.

\paragraph{Performance variation and executor dependence.}
Independent creations from the same model can still differ sharply.
The clearest example is an Opus Code harness that performs well under
\selfE{} but nearly collapses under Gemini because it hard-codes a
120-step limit around the original executor. Similar failures arise
from overly strict stopping rules in Data. A single generated harness
is therefore not representative, which motivates reporting avg@3.

\begin{figure}[htbp]
\centering
\includegraphics[width=\linewidth]{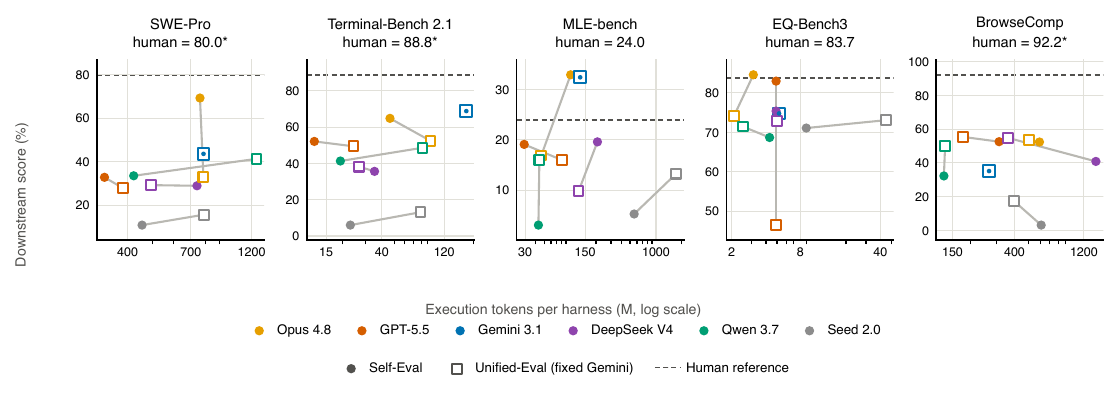}
\caption{\textbf{Cost vs.\ performance (RQ1).}
Downstream score against execution tokens (log scale) under \selfE{}
(filled circles) and \unifE{} (open squares; fixed executor
Gemini~3.1~Pro); dashed lines mark the human-engineered reference where
measured. All points are the avg@3 entries of Table~\ref{tab:rq1self} and
Table~\ref{tab:rq1unified}; Gemini is the fixed-executor control, so its
two markers coincide and appear as a dot inside a square. Similar quality
can differ by close to an order of magnitude in cost---on MLE-bench,
GPT-5.5 reaches a medal rate of 19.1 with 29.3M tokens while DeepSeek~V4
reaches 19.6 with 208.4M. We report this performance--token trade-off
directly rather than collapsing it into a single cost-adjusted score
(Section~\ref{sec:metrics}).}
\label{fig:cost}
\end{figure}

Fixing the executor changes the ranking substantially. Qwen, Seed, and
DeepSeek improve in several Data and Search settings under Gemini,
whereas Opus and GPT-5.5 are often stronger with their own executors.
Thus, \selfE{} reflects harness design, executor capability, and the
compatibility between them. Cost is similarly uneven: MLE-bench token
use varies by about nineteen-fold, yet higher cost does not reliably
produce a higher score (Figure~\ref{fig:cost}).
\begin{table}[H]
\centering
\scriptsize
\setlength{\tabcolsep}{2.8pt}
\renewcommand{\arraystretch}{1.05}
\resizebox{\linewidth}{!}{%
\begin{tabular}{l D{.}{.}{-1}D{.}{.}{-1} D{.}{.}{-1}D{.}{.}{-1} D{.}{.}{-1}D{.}{.}{-1} D{.}{.}{-1}D{.}{.}{-1} D{.}{.}{-1}D{.}{.}{-1} D{.}{.}{-1}}
\toprule
 & \multicolumn{2}{c}{\textbf{SWE-Pro}}
 & \multicolumn{2}{c}{\textbf{Term.-2.1}}
 & \multicolumn{2}{c}{\textbf{MLE-bench}}
 & \multicolumn{2}{c}{\textbf{EQ-Bench3}}
 & \multicolumn{2}{c}{\textbf{BrowseComp}}
 & \textbf{Avg.} \\
\cmidrule(lr){2-3}\cmidrule(lr){4-5}\cmidrule(lr){6-7}\cmidrule(lr){8-9}\cmidrule(lr){10-11}
\textbf{Creator (\selfE{})} & \multicolumn{1}{c}{succ.} & \multicolumn{1}{c}{tok.} & \multicolumn{1}{c}{acc.} & \multicolumn{1}{c}{tok.}
 & \multicolumn{1}{c}{medal} & \multicolumn{1}{c}{tok.} & \multicolumn{1}{c}{score} & \multicolumn{1}{c}{tok.} & \multicolumn{1}{c}{acc.} & \multicolumn{1}{c}{tok.} & \multicolumn{1}{c}{score} \\
\midrule
\rowcolor{lightgray}
\emph{Seed harness $H_{\mathrm{seed}}$} & 0.0 & \multicolumn{1}{c}{---} & 0.0 & \multicolumn{1}{c}{---} & 0.0 & \multicolumn{1}{c}{---} & 0.0 & \multicolumn{1}{c}{---} & 0.0 & \multicolumn{1}{c}{---} & 0.0 \\
\midrule
\rowcolors{1}{tabrow}{white}
Opus 4.8 High & 69.3 & 759.5 & 64.8 & 46.2 & 32.9 & 101.7 & 84.6 & 3.1 & 52.4 & 593.9 & 67.8 \\
GPT-5.5 High & 32.8 & 325.6 & 52.1 & 12.2 & 19.1 & 29.3 & 83.0 & 4.9 & 52.6 & 313.5 & 55.1 \\
Gemini 3.1 Pro High & 43.6 & 783.6 & 68.8 & 177.1 & 32.4 & 131.3 & 74.8 & 5.2 & 35.2 & 267.4 & 55.6 \\
DeepSeek V4 Pro High & 28.9 & 739.7 & 35.6 & 35.2 & 19.6 & 208.4 & 75.4 & 4.9 & 40.9 & 1{,}449.9 & 45.2 \\
Qwen 3.7 Max & 33.5 & 421.9 & 41.3 & 19.3 & 3.1 & 42.9 & 68.7 & 4.3 & 32.3 & 130.9 & 44.0 \\
Seed 2.0 Pro High & 10.8 & 454.4 & 6.0 & 23.0 & 5.3 & 557.1 & 71.1 & 9.0 & 3.2 & 610.2 & 22.8 \\
\midrule
\emph{Human harness + paired model} & 80.0\textsuperscript{*} & \multicolumn{1}{c}{---} & 88.8\textsuperscript{*} & \multicolumn{1}{c}{---} & 24.0 & 66.5 & 83.7 & 9.2 & 92.2\textsuperscript{*} & \multicolumn{1}{c}{---} & 86.2 \\
\bottomrule
\end{tabular}}
\caption{\textbf{RQ1 / Harness Creation under \selfE{}.}
Unless noted otherwise, each creator independently builds and evaluates three harnesses and the score is avg@3. Each benchmark uses its native metric, \emph{tok.} is the mean execution tokens per harness in millions, and \emph{Avg.} is the unweighted mean over SWE-Pro, Terminal-Bench, EQ-Bench3, and BrowseComp. MLE-bench covers 33 physical cells and 2{,}475 results. The human row is a system-level reference; \textsuperscript{*} marks external results that we did not re-run in this experiment, and --- marks unavailable entries.}
\label{tab:rq1self}
\end{table}

\begin{table}[H]
\centering
\scriptsize
\setlength{\tabcolsep}{2.8pt}
\renewcommand{\arraystretch}{1.05}
\resizebox{\linewidth}{!}{%
\begin{tabular}{l D{.}{.}{-1}D{.}{.}{-1} D{.}{.}{-1}D{.}{.}{-1} D{.}{.}{-1}D{.}{.}{-1} D{.}{.}{-1}D{.}{.}{-1} D{.}{.}{-1}D{.}{.}{-1} D{.}{.}{-1}}
\toprule
 & \multicolumn{2}{c}{\textbf{SWE-Pro}}
 & \multicolumn{2}{c}{\textbf{Term.-2.1}}
 & \multicolumn{2}{c}{\textbf{MLE-bench}}
 & \multicolumn{2}{c}{\textbf{EQ-Bench3}}
 & \multicolumn{2}{c}{\textbf{BrowseComp}}
 & \textbf{Avg.} \\
\cmidrule(lr){2-3}\cmidrule(lr){4-5}\cmidrule(lr){6-7}\cmidrule(lr){8-9}\cmidrule(lr){10-11}
\textbf{Creator (\unifE{})} & \multicolumn{1}{c}{succ.} & \multicolumn{1}{c}{tok.} & \multicolumn{1}{c}{acc.} & \multicolumn{1}{c}{tok.}
 & \multicolumn{1}{c}{medal} & \multicolumn{1}{c}{tok.} & \multicolumn{1}{c}{score} & \multicolumn{1}{c}{tok.} & \multicolumn{1}{c}{acc.} & \multicolumn{1}{c}{tok.} & \multicolumn{1}{c}{score} \\
\midrule
\rowcolor{lightgray}
\emph{Seed harness $H_{\mathrm{seed}}$} & 0.0 & \multicolumn{1}{c}{---} & 0.0 & \multicolumn{1}{c}{---} & 0.0 & \multicolumn{1}{c}{---} & 0.0 & \multicolumn{1}{c}{---} & 0.0 & \multicolumn{1}{c}{---} & 0.0 \\
\midrule
\rowcolors{1}{tabrow}{white}
Opus 4.8 High & 33.0\textsuperscript{\ensuremath{\ddagger}} & 782.9 & 52.4 & 94.3 & 16.9 & 46.3 & 74.2 & 2.1 & 53.6 & 505.8 & 53.3\textsuperscript{\ensuremath{\ddagger}} \\
GPT-5.5 High & 27.8 & 383.9 & 49.4 & 24.3 & 16.0 & 80.5 & 46.5 & 4.9 & 55.4 & 177.8 & 44.8 \\
Gemini 3.1 Pro High & 43.6 & 783.6 & 68.8 & 177.1 & 32.4 & 131.3 & 74.8 & 5.2 & 35.2 & 267.4 & 55.6 \\
DeepSeek V4 Pro High & 29.2\textsuperscript{\ensuremath{\ddagger}} & 492.4 & 38.2\textsuperscript{\ensuremath{\ddagger}} & 26.8 & 9.8 & 125.5 & 72.9 & 5.0 & 54.8 & 362.8 & 48.8\textsuperscript{\ensuremath{\ddagger}} \\
Qwen 3.7 Max & 41.3 & 1{,}253.3 & 48.6 & 81.9 & 16.0 & 44.0 & 71.5 & 2.5 & 49.9 & 133.4 & 52.8 \\
Seed 2.0 Pro High & 15.6 & 786.8 & 13.1 & 79.0 & 13.3 & 1{,}717.5 & 73.1 & 44.9 & 17.3 & 398.4 & 29.8 \\
\bottomrule
\end{tabular}}
\caption{\textbf{RQ1 / Harness Creation under a fixed Gemini executor.}
Every creator harness is executed by Gemini~3.1~Pro. The Gemini row reuses the \selfE{} control and is not counted as a new physical cell. Every score entry is avg@3. Code cells marked \textsuperscript{\ensuremath{\ddagger}} contain one collapsed R3 replica; dropping it gives post-hoc clean sensitivity means of 49.1 for Opus on SWE-Pro and 43.8/57.3 for DeepSeek on SWE-Pro/Terminal-Bench. GPT-5.5's EQ-Bench3 cell is avg@3 (46.5); excluding its zero-valued first harness the mean is 69.7. All other definitions follow Table~\ref{tab:rq1self}.}
\label{tab:rq1unified}
\end{table}

\paragraph{Implementation behavior.}
The six creators follow distinct implementation strategies. Opus often
rewrites the execution stack; GPT-5.5 adds a large monolithic agent;
DeepSeek, Qwen, and Seed extend the seed with agent, tool, context, and
state modules; and Gemini mostly edits the runner in place. The 18 Code
artifacts add 17,111 net lines in total
(Table~\ref{tab:rq1-code-diff}), but edit size does not predict
performance. Gemini adds the fewest lines (1,006) yet obtains the best
Terminal-Bench score (68.8), suggesting that focused changes and
frequent verification matter more than code volume.

\begin{table}[t]
\centering
\small
\setlength{\tabcolsep}{4.2pt}
\renewcommand{\arraystretch}{1.06}
\begin{tabular}{lrrrrD{.}{.}{-1}D{.}{.}{-1}}
\toprule
\textbf{Creator} & \textbf{Files add/chg/del} & \textbf{Total net LOC}
& \textbf{Median/replica} & \textbf{Range} & \multicolumn{1}{c}{\textbf{SWE-Pro}} & \multicolumn{1}{c}{\textbf{Term.-2.1}} \\
\midrule
\rowcolors{1}{tabrow}{white}
Opus 4.8        & 19/7/16 & 2,470 & 698   & 656--1,116 & 69.3 & 64.8 \\
GPT-5.5         & 3/10/0  & 3,537 & 1,231 & 1,059--1,247 & 32.8 & 52.1 \\
Gemini 3.1 Pro  & 4/4/0   & 1,006 & 324   & 270--412 & 43.6 & 68.8 \\
DeepSeek V4 Pro & 14/4/1  & 3,242 & 988   & 931--1,323 & 28.9 & 35.6 \\
Qwen 3.7 Max    & 12/6/0  & 3,562 & 1,339 & 551--1,672 & 33.5 & 41.3 \\
Seed 2.0 Pro    & 14/6/0  & 3,294 & 1,200 & 868--1,226 & 10.8 & 6.0 \\
\bottomrule
\end{tabular}
\caption{\textbf{Edit size of the frozen RQ1 Code artifacts.}
File counts and total net LOC are summed over each creator's three independently created artifacts; the median and range are per artifact. The statistics cover only \texttt{harness/}, which is the creator's responsibility, and exclude the runtime substrate injected by the runner. Performance is \selfE{} avg@3.}
\label{tab:rq1-code-diff}
\end{table}

\begin{figure}[H]
  \centering
  \includegraphics[width=0.90\linewidth]{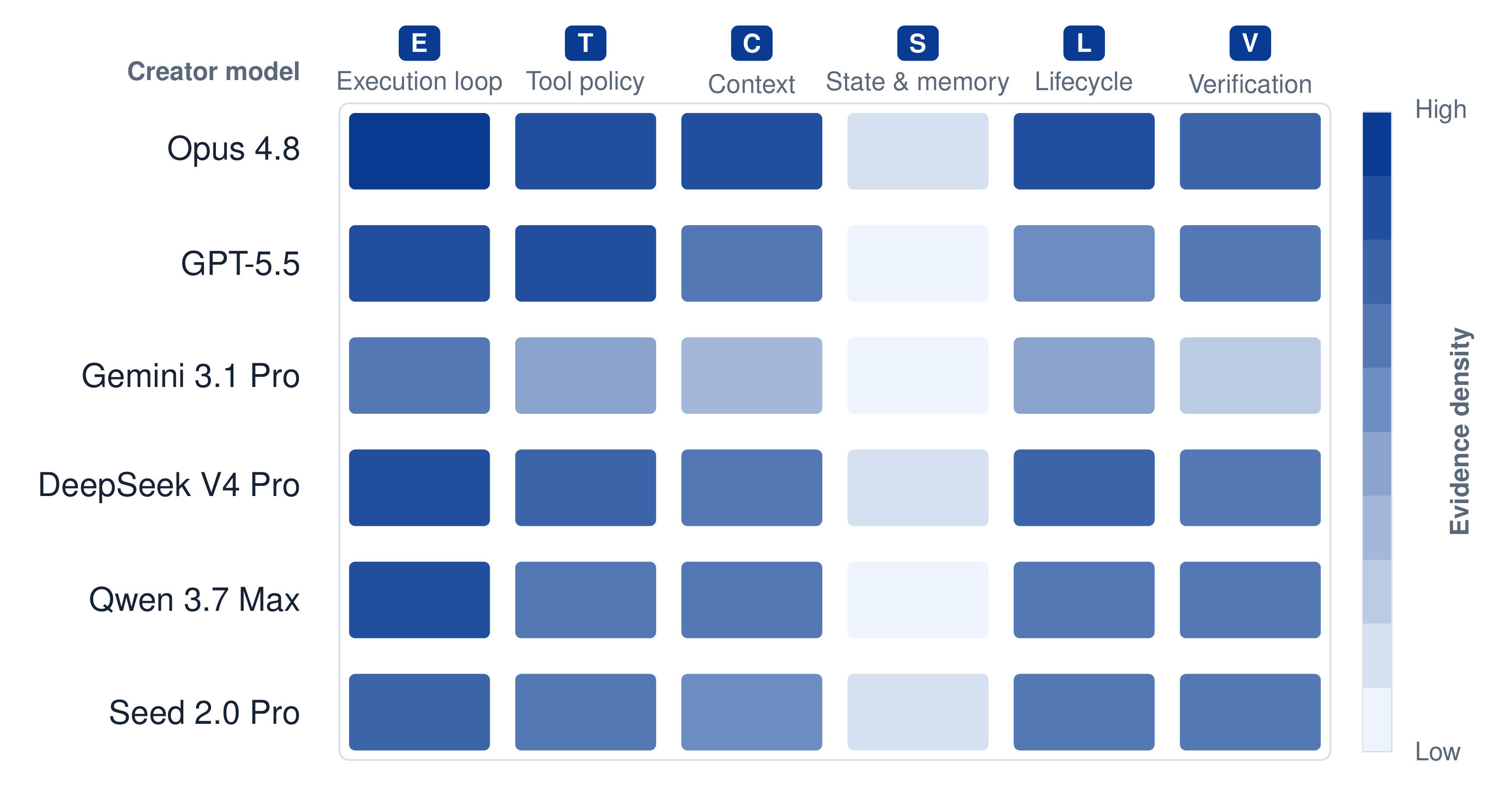}
  \caption{\textbf{Harness architecture evidence across fields and stages.}
  The six columns are execution loop, tool policy, context management, state and memory, lifecycle and recovery, and result verification. Each field is first normalized by the number of independent harnesses actually available and then aggregated over RQ1 Code/Data/Writing/Search and RQ2 Main; Seed has no RQ2 trajectory and is therefore built only from the four RQ1 fields. Full weight requires a mechanism to enter the main path or to trigger during a formal run, while declared code, configuration, or transient state receives only partial weight. Color shows mechanism evidence density, not score, significance, or causal effect.}
  \label{fig:architecture-coverage}
\end{figure}

All 18 Code harnesses implement an explicit execution loop; tools,
lifecycle control, and verification are complete in 13/18, 13/18, and
15/18 artifacts, respectively
(Figure~\ref{fig:architecture-coverage}). State and memory are the
clearest gap: 11/18 artifacts define a State class, but only one exposes
a state-saving interface and only one implements periodic checkpointing.
No checkpoint event appears in 26,679 recorded task trajectories.
Another recurring weakness is executor-specific configuration:
hard-coded step or output limits can make a functional harness fail when
the executor changes. Validation is also mostly syntactic; 441 of 2,325
executed Data tasks produce degenerate submissions that no harness
detects.

Some generated mechanisms never affect execution. Of 108 component
instances in Code, 72 trigger in real runs, 18 have only partial
evidence, and 18 are never observed; all unobserved instances concern
state and memory. The same pattern appears beyond Code: 124 of 587
Writing features are confirmed dead code, and 36 Data mechanisms sit on
dead paths. Small implementation errors can also disable an entire tool
chain. Self-test count alone is a weak signal: its Spearman correlation
with downstream score is only 0.13--0.26 and is not significant,
whereas revision calls reach 0.57 ($p\le.0005$). Testing helps when the
creator reads the failure, makes a targeted change, and re-verifies it.

\clearpage

\begin{figure}[H]
  \centering
  \includegraphics[width=\linewidth]{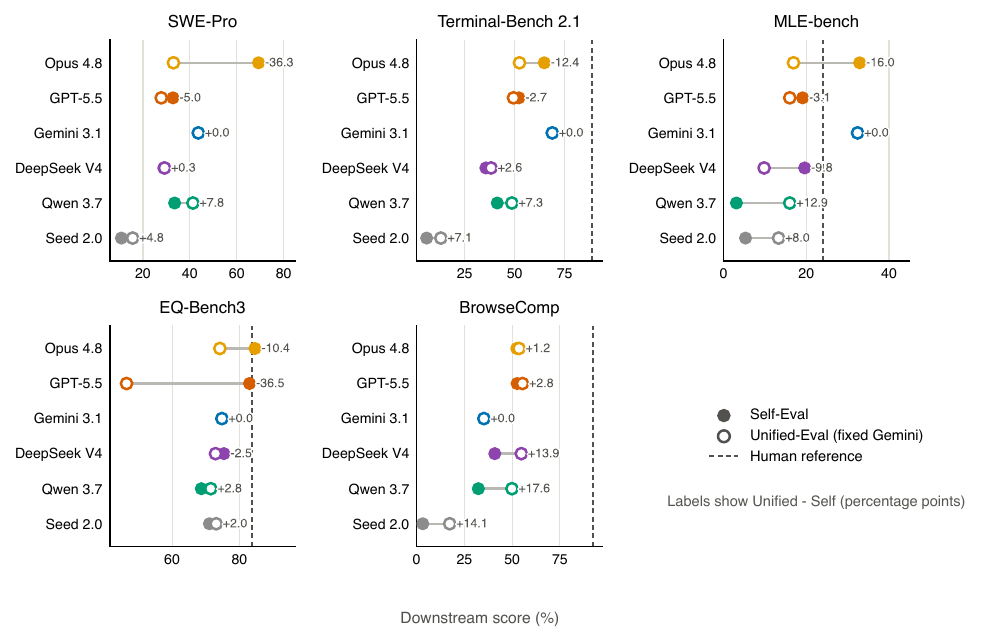}
  \caption{\textbf{Harness portability under a fixed Gemini executor.}
  Filled markers are \selfE{}, hollow markers are the fixed Gemini result for the same creator harness, labels are \unifE{} minus \selfE{}, and dashed lines are external human-engineered system references. All panels use the avg@3 values of Table~\ref{tab:rq1self} and Table~\ref{tab:rq1unified}; the clean sensitivity means for the two collapsed code cells are given in the caption of Table~\ref{tab:rq1unified}.}
  \label{fig:rq1-transfer}
\end{figure}

\paragraph{Executor transfer.}
Portability depends on the individual harness
(Figure~\ref{fig:rq1-transfer}). Several Qwen and
DeepSeek harnesses improve under Gemini, indicating that their original
executors were a bottleneck; Qwen gains 17.6 points on BrowseComp and
12.9 on MLE-bench. Opus shows the opposite pattern: its \selfE{}
SWE-Pro score falls from 69.3 to 33.0 under Gemini, and its Writing score
falls from 84.6 to 74.2. In the Opus Search harness, the duplicate-query
rate rises from $10.1\%$ to $88.2\%$ after the executor changes, showing
that its deduplication, review, and termination rules are adapted to the
original model. A runnable harness can therefore be used by another
model, but capability transfers only when its prompts, tool protocol,
budgets, and stopping rules remain compatible.

\clearpage

\FloatBarrier
\subsection{Harness Evolution}
\label{sec:exp-rq2}

\paragraph{Overall findings.}
All five self-runtime creators improve on the visible feedback pair,
but the gains shrink on held-out tasks. Opus~4.8 has the largest
held-out improvement at $+4.44$ points. Transfer is weaker under the
fixed Gemini executor: only Opus improves on held-out tasks, while the
other three lineages regress. Evolution can therefore produce useful
local changes, but the gains remain small and often specialize to the
current executor or feedback set.

\paragraph{Evaluation protocol.}
RQ2 asks whether a creator can improve its RQ1 Code harness using
downstream execution feedback. Tasks repeatedly evaluated during
Evolution form the \emph{feedback set}; tasks evaluated only after
Evolution, with results never returned to the creator, form the
\emph{held-out set}. We report the individual benchmark scores and an
equally weighted pair score:
\begin{equation}
  \bar{P}_{t}=\tfrac{1}{2}\left(P^{\mathrm{SWE100}}_{t}+P^{\mathrm{Term89}}_{t}\right),
\end{equation}
where the two terms are percentage scores and $t$ indexes the frozen
versions that completed a formal evaluation.

\begin{figure}[t]
\centering
\includegraphics[width=\linewidth]{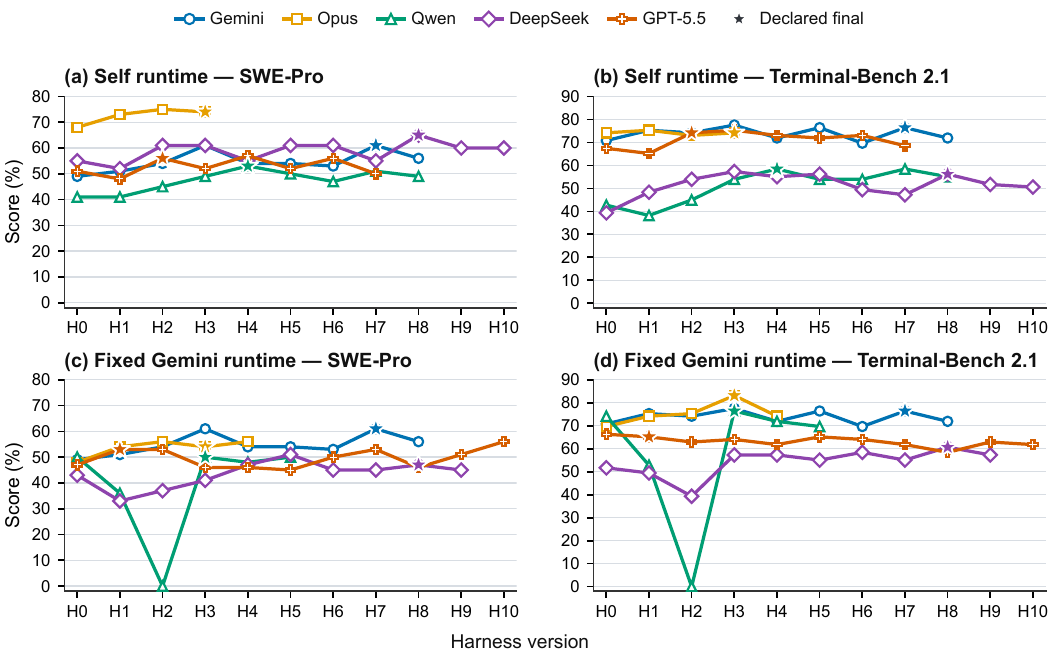}
\caption{\textbf{Feedback-set evolution trajectories on SWE-Pro and Terminal-Bench.}
The top row is self runtime and the bottom row is the fixed Gemini runtime; the left column is SWE-Pro-100 and the right column is Terminal-Bench-89. $H_0$ is the frozen RQ1 harness, $H_i$ are later frozen commits that completed a formal evaluation on both benchmarks, and stars mark the version declared by the creator. The Gemini control is reused in both settings, and each creator--runtime cell has a single trajectory.}
\label{fig:rq2-v6-trajectories}
\end{figure}

Each lineage starts from its RQ1 harness $H_0$. We run five
self-runtime trajectories and four fixed-Gemini ablations, using the
same creator, development environment, and starting harness. As defined
in Section~\ref{sec:settings}, an official version must complete both
the 100-task SWE-Pro and 89-task Terminal-Bench evaluations; probes are
diagnostic only. The nine lineages produce 73 official versions and 64
adjacent version switches; Figure~\ref{fig:rq2-v6-trajectories} shows
every feedback-set trajectory.

\begin{figure}[t]
\centering
\includegraphics[width=\linewidth]{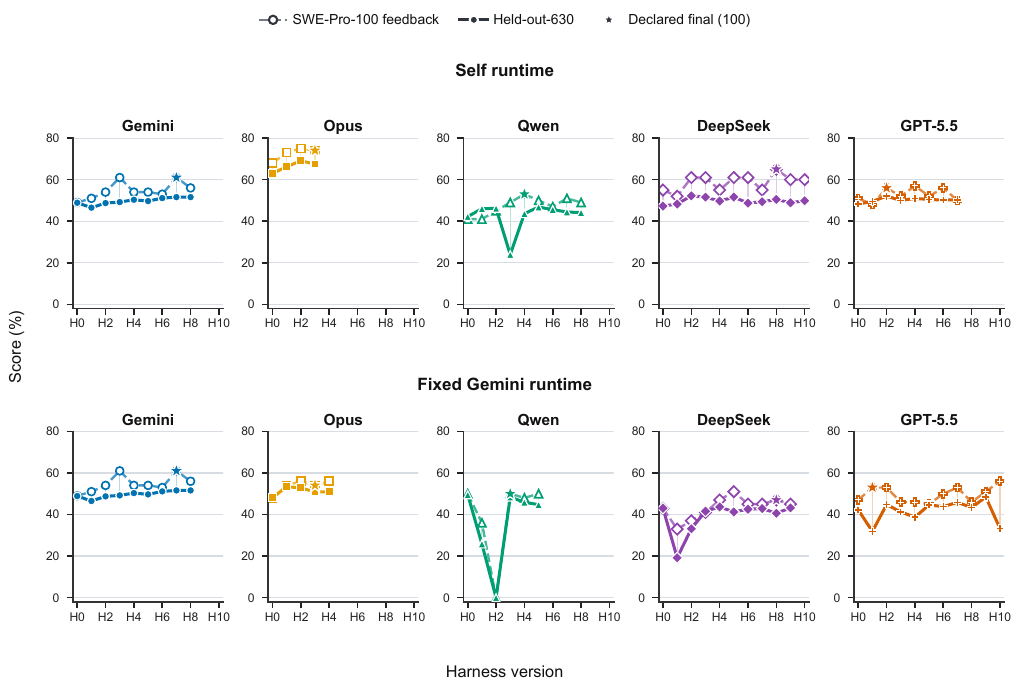}
\caption{\textbf{SWE-Pro-100 feedback trajectories and post-freeze held-out-630 performance.}
Each column isolates one creator model and overlays its visible 100-task
feedback trajectory with its post-freeze 630-task held-out trajectory.
The top row is self runtime and the bottom row is fixed Gemini. Stars
mark the final versions declared from 100-task feedback. Faint vertical
segments connect matched harness versions. The Gemini control is repeated
only for visual comparison.}
\label{fig:rq2-heldout}
\end{figure}

After all trajectories end, we evaluate every official version on the
630-instance SWE-Pro held-out split, drawn from the public-split instances
disjoint from the 100-task feedback set (Section~\ref{sec:settings}). These
scores are never shown to the creator and cannot
affect editing, stopping, or final version selection.
Table~\ref{tab:rq2-feedback-heldout} therefore
separates visible feedback gains from held-out generalization, and
Figure~\ref{fig:rq2-heldout} overlays the two trajectories for every
lineage.

\begin{table}[H]
\centering
\footnotesize
\setlength{\tabcolsep}{5.0pt}
\renewcommand{\arraystretch}{1.12}
\begin{tabular}{@{}llccr@{}}
\toprule
\textbf{Setting} & \textbf{Creator}
& \shortstack{\textbf{Feedback pair}\\$H_0\!\rightarrow H_{\mathrm{dec}}$}
& \shortstack{\textbf{Held-out-630}\\$H_0\!\rightarrow H_{\mathrm{dec}}$}
& \shortstack{\textbf{Held-out}\\\textbf{final gap}} \\
\midrule
\rowcolors{1}{tabrow}{white}
Self & Gemini 3.1 Pro & \shortstack{$59.9\!\rightarrow\!68.7$\\$(+8.8)$} & \shortstack{$48.89\!\rightarrow\!51.59$\\$(+2.70)$} & 0.00 \\
Self & Opus 4.8 & \shortstack{$71.1\!\rightarrow\!74.1$\\$(+3.0)$} & \shortstack{$63.02\!\rightarrow\!67.46$\\$(+4.44)$} & 1.59 \\
Self & Qwen 3.7 Max & \shortstack{$41.8\!\rightarrow\!55.7$\\$(+13.9)$} & \shortstack{$42.22\!\rightarrow\!43.65$\\$(+1.43)$} & 3.17 \\
Self & DeepSeek V4 Pro & \shortstack{$47.2\!\rightarrow\!60.6$\\$(+13.4)$} & \shortstack{$47.30\!\rightarrow\!50.48$\\$(+3.17)$} & 1.75 \\
Self & GPT-5.5 & \shortstack{$59.2\!\rightarrow\!65.1$\\$(+5.9)$} & \shortstack{$48.25\!\rightarrow\!52.06$\\$(+3.81)$} & 0.00 \\
\midrule
Fixed Gemini & Opus 4.8 & \shortstack{$58.8\!\rightarrow\!68.6$\\$(+9.7)$} & \shortstack{$48.10\!\rightarrow\!50.79$\\$(+2.70)$} & 2.54 \\
Fixed Gemini & Qwen 3.7 Max & \shortstack{$62.1\!\rightarrow\!63.2$\\$(+1.1)$} & \shortstack{$49.52\!\rightarrow\!48.41$\\$(-1.11)$} & 1.11 \\
Fixed Gemini & DeepSeek V4 Pro & \shortstack{$47.3\!\rightarrow\!53.8$\\$(+6.5)$} & \shortstack{$43.02\!\rightarrow\!40.63$\\$(-2.38)$} & 3.02 \\
Fixed Gemini & GPT-5.5 & \shortstack{$56.6\!\rightarrow\!59.1$\\$(+2.4)$} & \shortstack{$42.22\!\rightarrow\!31.90$\\$(-10.32)$} & 16.51 \\
\bottomrule
\end{tabular}
\caption{\textbf{RQ2 feedback-set gains and post-freeze held-out-630 generalization.}
The feedback pair is the unweighted mean of the SWE-Pro-100 and Terminal-Bench-89 percentage scores; the held-out columns use only the SWE-Pro-630 tasks run afterwards. The final gap is the best held-out score of a lineage minus the score of its declared version. The Gemini control is listed once.}
\label{tab:rq2-feedback-heldout}
\end{table}

\paragraph{Editing and feedback use.}
The median declared version changes eight files, adding 476 lines and
deleting 38 (Table~\ref{tab:rq2-change-summary}). Across 64 official switches, 58 change execution or control
flow, 37 change tools, 17 change lifecycle recovery, 16 change context,
and only four change state; no switch modifies a standalone verifier.
Edit size does not reliably predict improvement, and the same creator
may adopt different strategies under different executors. DeepSeek, for
example, expands its self-runtime harness but later rolls back much of a
fixed-Gemini rewrite after context compression breaks tool-message
pairing. Evolution therefore resembles local program search around
runtime feedback, where deletion can be as useful as addition.

\begin{table}[H]
\centering
\scriptsize
\setlength{\tabcolsep}{4.0pt}
\renewcommand{\arraystretch}{1.08}
\resizebox{\linewidth}{!}{%
\begin{tabular}{llccl}
\toprule
\textbf{Setting} & \textbf{Creator} & \textbf{Official switches}
& \textbf{$H_0\!\rightarrow H_{\mathrm{dec}}$ cumulative diff} & \textbf{Main edit focus} \\
\midrule
\rowcolors{1}{tabrow}{white}
Self & Gemini 3.1 Pro & 8  & 11 files, $+692/-38$ & Tool output, editing, timeouts, patch cleanup \\
Self & Opus 4.8       & 3  & 8 files, $+450/-18$ & Completion gate, self-review, process recovery \\
Self & Qwen 3.7 Max   & 8  & 9 files, $+1101/-106$ & Empty-response/parse recovery, auto-submit checkpoints \\
Self & DeepSeek V4 Pro& 10 & 9 files, $+1730/-58$ & Completion logic, tool extensions, workspace pre-analysis \\
Self & GPT-5.5        & 7  & 4 files, $+496/-49$ & Final-state review gate and artifact tracking \\
Fixed Gemini & Opus 4.8       & 4  & 3 files, $+227/-21$ & Pre-completion verification, scratch-file cleanup \\
Fixed Gemini & Qwen 3.7 Max   & 5  & 4 files, $+219/-96$ & Context and message-sanitizer rewrite \\
Fixed Gemini & DeepSeek V4 Pro& 9  & 8 files, $+476/-12$ & Message-protocol recovery, workspace discovery \\
Fixed Gemini & GPT-5.5        & 10 & 6 files, $+297/-34$ & Probes, branch rollback, final-state review \\
\bottomrule
\end{tabular}}
\caption{\textbf{Code edit size of the nine RQ2 lineages.}
``Official switches'' counts only adjacent official versions; the cumulative diff compares $H_0$ with the version declared by the creator for each lineage. The edit focus describes behavior observed in that single trajectory and is not a general property of the corresponding creator model.}
\label{tab:rq2-change-summary}
\end{table}

Eight of the nine lineages complete at least one full loop from reading
results to editing, re-evaluation, and version selection. Failure
diagnosis remains the weakest step: the dedicated trajectory interface
is called only twice, and explicitly inspected cases cover just
$0.5\%$--$40.2\%$ of the 189 feedback tasks, depending on the lineage.
Creators instead rely on custom scripts and small probes, both of which
can disagree with the full evaluation; one GPT-5.5 candidate passes all
five Terminal probes but scores only 0.584 on the full set. Opus gives
the clearest positive example: it finds that 99 of 100 runs report
success while only 48 pass, traces the gap to premature completion, and
adds a completion check. Feedback is most useful when it exposes a
concrete failure mode and the resulting change is verified end to end.

\paragraph{Stability and final-version selection.}
Evolution is not monotonic. Of the 64 official switches, eight regress
on both benchmarks, 16 show a single-benchmark regression, three show
a cross-benchmark trade-off, seven produce no measurable change, 27
report gains that remain inside the repeated-run noise band, two have
clear positive evidence beyond the noise band, and one contains no
executable code change. The same commit can vary by about $\pm4.75$ pair-score points,
so small gains cannot be attributed to code changes from score alone.
Added code is not necessarily active either: of 169 new functions or
classes, 113 are reachable from the entry point, 31 are reachable only
through dead code, and 25 have no caller. Opus's completion gate is a
positive example with path and case-level evidence; Qwen's message
sanitizer is the opposite, breaking valid Gemini tool-result sequences.

Creators usually select a version near the best visible feedback score,
but that choice rarely matches the best held-out version. All five
self-runtime declarations improve over $H_0$ on held-out tasks, with
gains of $+1.43$ to $+4.44$ points and a mean gain of $+3.11$. Under
fixed Gemini, however, only Opus improves and the other three regress.
Across 64 comparable switches, feedback and held-out scores move in the
same direction only 34 times ($53.1\%$), and only 2/9 declared versions
are held-out optimal. Visible feedback is therefore useful for local
search but unreliable for final selection: repeatedly optimizing a
noisy score can favor a lucky run and amplify overfitting.

\FloatBarrier

\section{Related Work}
\label{sec:related}

\harness lies at the intersection of agent benchmarking, autonomous
agent construction, and harness evolution.

\paragraph{Agent benchmarks and harness development.}
Benchmarks such as SWE-bench~\citep{ICLR2024_edac78c3},
GAIA~\citep{ICLR2024_25ae35b5}, WebArena~\citep{ICLR2024_4410c071},
$\tau$-bench~\citep{yao2024taubenchbenchmarktoolagentuserinteraction},
and AgentBench~\citep{ICLR2024_e9df36b2} standardize tasks,
environments, and scoring, but generally evaluate task execution under
a selected harness. Harness-Bench~\citep{yao2026harness} instead
measures how harness choice changes model performance. The Meta-Agent
Challenge~\citep{lu2026meta} directly evaluates development: a
meta-agent iteratively programs an agent artifact in a sandbox and is
scored on protected held-out tests across five domains. It is closely
related to \hC{}, while \harness also studies continued development,
separates creator and executor models, and measures execution cost.

HarnessOpt-Bench~\citep{ursekar2026harnessopt} is the closest concurrent
benchmark to \hE{}. An LLM optimizer receives a seed harness, graded
feedback, and a fixed evaluation budget; a trusted environment then
scores its nominated candidate by normalized gain on an inaccessible
test partition. Its focus is optimizing a provided harness, although
its near-empty GAIA seed also requires construction. In contrast,
\harness connects from-scratch Creation to Evolution and evaluates the
same frozen artifact under self and fixed runtime models. It uses
in-trajectory scores as feedback, then tests every frozen version on a
disjoint 630-task SWE-Pro set, separating adaptation from held-out
generalization throughout the trajectory.

Evo-Bench~\citep{huang2026evobenchlanguagemodelsimprove} also evaluates
\hE{} by asking evolver models to improve a shared CodeAct seed while
holding the runtime model fixed. It scores each lineage's final revision
on a disjoint, sensitivity-calibrated multi-domain suite. \harness
instead connects Creation and Evolution, includes both self- and
fixed-runtime views, measures execution-token cost, evaluates transfer
to another executor, and scores every frozen version on its held-out
split. Thus, Evo-Bench emphasizes cross-domain final-revision quality,
whereas \harness also studies how held-out performance changes along a
feedback-driven trajectory.

\paragraph{Automated agent design and evolution.}
Human-engineered systems such as Claude Code~\citep{anthropic2024claude_code},
Codex~\citep{openai2025codex}, OpenHands~\citep{wang2025openhandsopenplatformai},
and SWE-agent~\citep{NEURIPS2024_5a7c9475} combine execution loops,
tools, context management, and recovery mechanisms
\citep{rombaut2026insidescaffoldsourcecodetaxonomy}. ADAS~\citep{ICLR2025_36b7acf6},
AFlow~\citep{ICLR2025_5492ecbc}, MASS~\citep{zhou2026multiagentdesignoptimizingagents},
and EvoAgentX~\citep{wang-etal-2025-evoagentx} search over prompts,
workflows, operators, or agent topologies; other systems let models
construct more of the agent in natural language or code
\citep{pan2026natural_language_harnesses,ning2026code_agent_harness,
li2026opensageselfprogrammingagentgeneration}.

Recent systems optimize executable harness code directly. VeRO
\citep{ursekar2026vero} provides versioned snapshots, budget-controlled
evaluation, and structured traces. Meta-Harness~\citep{lee2026metaharness}
uses a coding-agent proposer that can inspect prior candidates, scores,
and traces before proposing the next harness. Other methods improve
prompts, workflows, memory, skills, tools, or code between episodes
\citep{shinn2023reflexionlanguageagentsverbal,
wang2023voyageropenendedembodiedagent,liu2025sew,tao2026sepo,
he2025evotest,luo2026self,karten2026continual,chen2026harnessx,
lin2026agentic,lee2026recursiveharnessselfimprovement}.
Self-Harness~\citep{zhang2026self} derives model-specific edits from
failures and applies regression testing; HarnessFix~\citep{chen2026harnessfix}
uses a provenance- and control-flow-aware representation for localized
repair; DemoEvolve~\citep{che2026demoevolve} uses demonstrations when
reward feedback is unreliable; and HarnessCompass~\citep{zhang2026harnesscompass}
addresses overfitting and component interference. Harness-R1
\citep{shao2026harnessr1} instead trains a harness engineer to convert
failure batches into validated patches. These works propose methods or
infrastructure for harness improvement; \harness evaluates how well
general-purpose frontier models perform the broader developer role
under one Creation-and-Evolution protocol.

\paragraph{Evaluating harness evolution.}
Final task gain can conflate informed diagnosis with blind search.
Priority-ranking evaluation~\citep{ong2026priority} tests whether an
optimizer identifies the components most worth changing. Harness
Updating Is Not Harness Benefit~\citep{selfharness} separates producing
a useful update from an executor's ability to exploit it, motivating
our creator--executor separation and \selfE{}/\unifE{} views. SEAGym
\citep{zheng2026seagym} records intermediate snapshots, cost, and
in- and out-of-distribution results, showing that later updates need not
preserve held-out gains. Matched-budget studies likewise find that
harness evolution can overfit its search benchmark and may not beat
simpler search baselines~\citep{wang2026rethinking}. \harness therefore
freezes runnable artifacts, records development trajectories and
execution cost, and evaluates transfer across runtime models. We treat
its Evolution trajectories as adaptation to feedback-bearing sets and
assess held-out generalization afterward on disjoint SWE-Pro tasks;
matched-search evaluation remains future work.

\section{Discussion, Limitations, and Conclusion}
\label{sec:discussion}

The results show why harness development should be evaluated directly.
Creation performance varies sharply by domain: under \selfE{}, current
models match the human reference in writing and exceed it in
machine-learning experimentation, but remain far behind in search and
research and still trail it in code. Cross-executor comparisons show that
some harnesses improve under a stronger executor, while others exhibit
creator co-adaptation. Evolution is harder still: useful intermediate
updates are often erased by later changes, and more updates do not
guarantee a positive final gain. Together, these findings separate the
quality of the persistent execution system from the capability of the
model running inside it. The fixed-Gemini Evolution ablation sharpens this
point: changing only the runtime binding can substantially move $H_0$ and
alter which harness changes are useful.

\subsection{Limitations}
\label{sec:disc-limits}
The four categories cover many but not all real deployments. Human
baselines are uneven and not guaranteed optimal. \unifE{} reduces but
cannot fully remove executor-model differences, since harness--model
interaction is complex. The behavioral comparisons are descriptive and
are limited by incomplete benchmark coverage. Evolution currently has
one trajectory per creator--runtime cell and one unfinished main-runtime
cell, and its post-freeze held-out evaluation covers SWE-Pro only, so
the trajectories do not support uncertainty estimates or
population-level comparisons. The development environment $D$ is held
fixed across both stages; whether an evolved harness can itself serve
as the development environment for further evolution is left to future
work. Finally,
\harness measures model-external learning and
does \emph{not} claim heuristic learning can replace parameter
training.

\subsection{Conclusion}
\label{sec:disc-conclusion}
\harness moves agent evaluation from \emph{whether a model can solve
tasks inside a fixed system} to \emph{whether it can create and
maintain the systems that solve future tasks}. Through a four-category
human baseline corpus, from-scratch creation tasks, feedback-driven
evolution, Self- and Unified-Eval, and comparisons with human-engineered
references, it makes agent-built
execution harnesses a measurable object. If model weights are one place
intelligence accumulates, the harness is another: explicit, inspectable,
testable, reusable, and continually improvable through failure,
feedback, and real engineering pressure.

\paragraph{Ethics statement.}
\harness is built from publicly available benchmark suites and
open-source harnesses; no human-generated content is collected.
Automated harness construction risks amplified insecure tool use, so the
benchmark states explicit constraints on what a submitted harness may do
and audits compliance after every run (Section~\ref{sec:eval}); no
violation was observed in this study, and we release the audit artifacts
so the check can be repeated. Creation runs and downstream benchmark
tasks execute in containers, but that boundary is provisioned for
reproducibility rather than for containment; anyone reusing the generated
harnesses should treat them as untrusted code and isolate them more
strictly than we did.

\clearpage
\section{Contributions}

\textbf{Core Contributors} \\
Yuhao Wu, Jingyuan Zhang, Jiajun Shi

\vspace{0.5em}

\textbf{Contributors} \\
Xinping Lei, Qingshui Gu, Yuxuan Zhang, Zexuan Wang, Chen He,
Chen Huang, Maojia Song, Zhiyuan Zeng, Shaowen Wang, Jinkai Liu, Yunfeng Shi, Jiaheng Liu

\vspace{0.5em}

\textbf{Corresponding Authors} \\
Yuhao Wu (\email{wuyuhao.621@bytedance.com})\\
Shen Yan (\email{sheny@bytedance.com})\\
Wenhao Huang (\email{huang.wenhao@bytedance.com})\\
Ge Zhang (\email{gezhang@umich.edu})\\
Wenxuan Zhang (\email{wxzhang@sutd.edu.sg})

\clearpage
\bibliographystyle{plainnat}
\bibliography{custom}

\clearpage
\beginappendix
\section{Candidate Harness Systems}
\label{sec:appcorpus}

The benchmark draws candidate systems from four categories. A system
may define the scope of a category, serve as a human-engineered
counterpart, supply an accessible update history, or act as a
development environment; inclusion below does not imply that every system serves
every role. The final pinned set records these role assignments along
with commit hashes and licenses where applicable and is released with
the benchmark.

\begin{itemize}[leftmargin=1.4em,itemsep=1pt]
  \item \textbf{Code agent:} Claude~Code, OpenCode, OpenHands,
        SWE-agent, mini-SWE-agent.
  \item \textbf{Notebook / data-analysis:} DataAgent, DB-GPT.
  \item \textbf{Writing agent:} AutoResearchClaw, webnovel-writer.
  \item \textbf{Research / retrieval:} Alibaba-NLP/DeepResearch,
        dzhng/deep-research, modelscope/ms-agent, gpt-researcher.
\end{itemize}

\section{Experimental Configuration}
\label{sec:apptables}

Table~\ref{tab:creatorconfig} records the creator LLM, development
environment, and decoding configuration used by the reported experiments. Sampling
parameters follow each provider's official defaults; all creators run
at high reasoning effort with streaming enabled, and output length is
set to the endpoint maximum. Participation is stage-specific, so the
presence of a creator in this table does not imply complete coverage of
every benchmark.

\begin{table}[htbp]
\centering
\small
\setlength{\tabcolsep}{5pt}
\renewcommand{\arraystretch}{1.05}
\begin{tabular}{l l c c c r}
\toprule
\textbf{Creator $L_C$} & \textbf{Development env. $D$} & \textbf{Temp.}
 & \textbf{Top-$p$} & \textbf{Top-$k$} & \textbf{Max output} \\
\midrule
Opus 4.8        & Claude Code 2.1.177 & 1.0 & default & default & 128{,}000 \\
GPT-5.5         & Codex 0.144.3       & default & default & --- & 128{,}000 \\
Gemini 3.1 Pro  & Claude Code 2.1.177 & 1.0 & 0.95 & 64 & 65{,}100 \\
DeepSeek V4 Pro & Claude Code 2.1.177 & 1.0 & 0.95 & --- & 131{,}072 \\
Qwen 3.7 Max    & Claude Code 2.1.177 & 0.6 & 0.95 & 20 & 65{,}536 \\
Seed 2.0 Pro    & Claude Code 2.1.177 & 1.0 & 0.70 & --- & 131{,}072 \\
\bottomrule
\end{tabular}
\caption{\textbf{Creator-LLM and development-environment configuration.}
\emph{default} means that the provider does not expose or we do not
override the value; --- means not applicable. Max output is measured in
tokens. The Opus~4.8 value is the measured endpoint cap.}
\label{tab:creatorconfig}
\end{table}

\subsection{Downstream execution configuration}
\label{sec:downstreamconfig}

Data-analysis (MLE-bench) downstream runs execute each task in an
isolated container with one NVIDIA A800-SXM4-80GB GPU (80\,GB of GPU
memory), 14 vCPUs, and 227\,GiB of RAM. Each task has a wall-clock
limit of 36{,}000\,s, split into a 34{,}200\,s budget for the generated
agent harness and a 1{,}800\,s reserve for the fixed grader, together
with a 500-step cap. Dataset download and preparation complete before
the task clock starts and do not consume this budget. The RQ2 code
benchmarks run each task with the same 500-step cap and a 7{,}200\,s
limit, as stated in the evolution contract of
Appendix~\ref{sec:appprompts}.

\subsection{Human-engineered reference systems}
\label{sec:humanrefs}

For each downstream benchmark, we use the highest publicly available
system-level result that we could verify from the benchmark's official
leaderboard or the corresponding system report. These references pair a
human-engineered harness with the executor model used by that system;
they are not scores obtained with one common executor. Table~\ref{tab:humanrefs}
records the exact pairs used in the main paper.

The three starred values in Table~\ref{tab:rq1self} are external
reports rather than local reruns: SWE-Pro 80.0 for Claude Fable~5,
Terminal-Bench~2.1 88.8 for GPT-5.6 Sol, and BrowseComp 92.2 for
GPT-5.6 Sol. Each value is taken from the corresponding benchmark row
in OpenAI's official GPT-5.6 release report~\citep{openai2026gpt56results}.

\begin{table}[htbp]
\centering
\small
\setlength{\tabcolsep}{5pt}
\renewcommand{\arraystretch}{1.05}
\begin{tabular}{l l l}
\toprule
\textbf{Benchmark} & \textbf{Human-engineered harness} & \textbf{Paired executor model} \\
\midrule
SWE-Pro & Public coding-agent setup & Claude Fable~5 \\
Terminal-Bench~2.1 & OpenAI agent setup & GPT-5.6 Sol \\
MLE-bench & MLEvolve & Gemini~3.1 \\
EQ-Bench3 & Kimi Writer & Opus~4.8 \\
BrowseComp & OpenAI browsing stack & GPT-5.6 Sol \\
\bottomrule
\end{tabular}
\caption{\textbf{Human-engineered reference systems.} Each row records
the harness--executor pair associated with the public system-level
reference used in Table~\ref{tab:rq1self}.}
\label{tab:humanrefs}
\end{table}

\subsection{Creation performance and execution cost}

\clearpage
\subsection{Evolution execution cost by frozen version}

\begin{figure}[htbp]
\centering
\includegraphics[width=\linewidth]{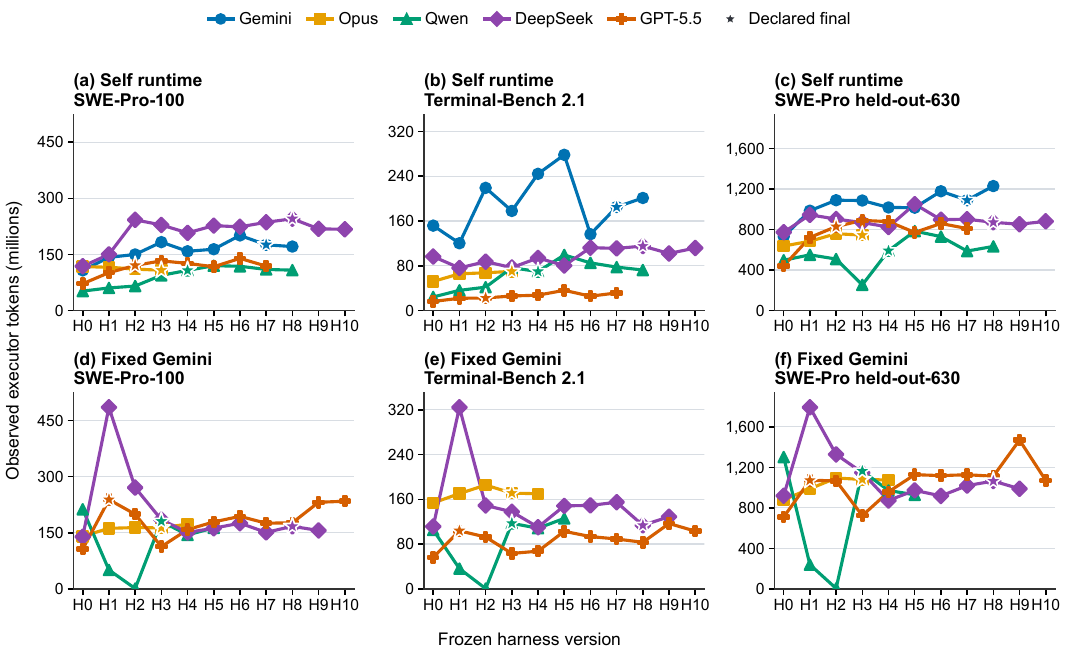}
\caption{\textbf{RQ2 executor-token cost across frozen harness versions.}
The top row uses self runtime and the bottom row uses the fixed Gemini
runtime; columns show SWE-Pro-100, Terminal-Bench~2.1, and the
post-freeze SWE-Pro held-out-630 split. Each point sums the observed
task-agent/runtime \texttt{total\_tokens} for one frozen harness and
benchmark leg. Creator, judge, and probe tokens are excluded.}
\label{fig:rq2-executor-token-cost}
\end{figure}

Figure~\ref{fig:rq2-executor-token-cost} complements the RQ2 score
trajectories with executor-side usage across every frozen harness
version. We keep task-agent/runtime usage separate from the tokens spent
by the creator, judge, and diagnostic probes.

\FloatBarrier

\section{Harness Interface Specification}
\label{sec:appinterface}

Each admissible harness implements six functional modules:

\begin{specbox}
\begin{verbatim}
execution.py   — run(task) -> Result
                 step(state, observation) -> Action
tools.py       — register(toolspec) -> None
                 call(name, **params) -> Observation
context.py     — build(task, history, state) -> Prompt
                 compress(messages) -> Messages
state.py       — save(checkpoint) -> None
                 load(id) -> State
                 resume() -> State
lifecycle.py   — beforeAction(action) -> Action | Abort
                 afterAction(action, result) -> None
                 onFailure(error) -> Recovery
                 onTimeout() -> Graceful
evaluation.py  — evaluate(result, criteria) -> Score
                 recordTrajectory(step) -> None
\end{verbatim}
\end{specbox}

\noindent All methods return JSON-serializable objects. The reference
seed implementation, the audit script, and the held-out task splits
will be released with the benchmark.

\subsection{Seed Harness Skeleton}
\label{sec:appseed}

The weak seed shared across Creation domains
(Section~\ref{sec:settings}) supplies only the runnable floor beneath
the contract above: a stable CLI, runtime model configuration, audit
writers, and policy-free primitives. Its concrete packaging may vary
with the execution environment, but no variant contains a
task-solving policy. Figures~\ref{fig:weak-seed-env}
and~\ref{fig:creator-contract} in the main text give the
domain-general structure and separate the shared audit contract from
domain-specific final artifacts.

\begin{specbox}
\begin{verbatim}
seed workspace/
  dev runner          # public development-feedback runner
  harness/            # the package: python -m harness
    entry module      # stable task/config/output CLI
    seed runner       # parse inputs; optional summary probe
    audit contract    # result.json · trajectory.jsonl ·
                      # response.md · stdout/stderr logs
    llm gateway       # configured connectivity, no policy
    primitives/       # passive helpers, no policy
      paths           #   resolve · contains · info
      files           #   read · write · replace · json io
      search          #   list tree · glob · grep
      process         #   run command
      artifact io     #   create · validate · record paths
\end{verbatim}
\end{specbox}

\noindent The unmodified seed performs one non-acting pass: it parses
the task and environment, optionally probes the configured LLM with a
summary-only prompt, and terminates with status \texttt{partial} after
writing the audit envelope. It provides no execution loop, tool
policy, context management, state or memory, failure recovery, or
verifier. The creator must therefore implement these modules using only
feedback from public development tasks; hidden tasks, answers, and
official scores remain withheld.

The interface and honest-status semantics are shared, but the
authoritative final artifact is domain-specific. Code requires real
repository changes and a patch; data analysis requires a
scorer-readable submission; writing requires final user-facing prose;
and research/search requires a concise
answer grounded in retrieved evidence. Each frozen evaluator reads
the corresponding authoritative artifact. Producing only the common
JSON and log files is therefore an incomplete execution and, as shown
in Table~\ref{tab:rq1self}, the unmodified seed scores zero across all
five downstream benchmarks.

\section{Evaluation Settings and Roles in Full}
\label{sec:appevalmetrics}

This appendix gives the evaluation-setting and model-role definitions
summarized in Section~\ref{sec:eval}.

\subsection{Self and unified evaluation}
\label{sec:appeval}

We evaluate a created harness under two executor settings. \selfE{}
asks whether the harness helps the model that built it, whereas
\unifE{} tests the same generated harnesses with a common executor.

Under \selfE{}, $L_E=L_C$: the creator LLM runs the hidden tasks using
its own harness $H$. This is the regime users actually deploy, and it
measures model--harness co-design: whether a model can build a harness
suited to its own capability boundary. Under \unifE{}, a single fixed
$L_E$ runs every harness produced by every $L_C$, removing executor-LLM
differences so far as possible; $L_E$ is held constant across all
comparisons (Gemini~3.1~Pro in the reported fixed-executor ablation).
The executor may itself appear as a creator; when included, its
own-harness cell then coincides with its \selfE{} run---and
cross-creator differences under the shared executor still isolate
harness quality from executor ability. A high unified score indicates the harness is a
transferable software asset rather than a fit to one model.
Human-engineered systems are external references, not
a third executor setting. They show the distance from selected mature
systems but are not paired controls under a common executor and should
not be interpreted as an absolute ceiling.

\subsection{Model roles}
\label{sec:approles}

Following Section~\ref{sec:taskfamilies}, $L_C$ builds or modifies the
harness, $D$ supplies file reading, code editing, testing, and debugging,
and $L_E$ runs downstream tasks only after $H$ is frozen. The evaluator
$J$ scores the resulting task output. Separating these roles prevents us
from attributing support from $D$ or execution ability from $L_E$ to the
quality of $H$. The reported experiments therefore record $D$ for each
creator configuration and fix $L_E$ and $J$ within every comparison.

\section{Representative System Prompts}
\label{sec:appprompts}

This appendix presents the two system prompts that define the
representative harness-development settings studied in RQ1 and RQ2.
Task-specific prompts and auxiliary workspace documents are omitted
because the purpose here is to illustrate the system-level contracts.
The RQ1 prompt was shared across creation domains and Creator models.
For the rendered RQ2 V6 prompt, only run-specific filesystem paths and
baseline evaluation identifiers are replaced by angle-bracketed
placeholders; all substantive instructions are unchanged.

\subsection{RQ1: shared harness-creation system prompt}

\begin{tcblisting}{
  enhanced,
  breakable,
  listing only,
  colback=specbg,
  colframe=medgray,
  boxrule=0.5pt,
  arc=1.5mm,
  borderline west={2pt}{0pt}{seedaccent},
  left=2mm,
  right=2mm,
  top=1mm,
  bottom=1mm,
  before skip=8pt,
  after skip=8pt,
  listing options={
    basicstyle=\ttfamily\scriptsize,
    breaklines=true,
    breakatwhitespace=false,
    columns=fullflexible,
    keepspaces=true,
    showstringspaces=false,
    literate={—}{{---}}1
  }
}
# System Prompt: Open Harness Construction Contract

You are an agent harness engineer. Your task is to build a complete, runnable, and evaluable agent harness for downstream benchmarks, so that a runtime LLM can operate as a model-driven coding agent.

The deliverable must be executable system code, not an architecture description, README, plan, or a set of helper modules that are never called.

---
## Terminology

To avoid ambiguity, this task uses the following terms:

- Runtime LLM: the model called by the harness during downstream task execution.
- Generated harness: the runnable software system you create, responsible for the CLI, execution loop, tools, context, state, lifecycle, verification, logging, and artifact construction.
- Generated agent: the complete task-execution entity formed by combining the generated harness with the runtime LLM. In other words, generated agent = generated harness + runtime LLM.
- Creator: the model currently building the harness.
- Metaharness: the outer workbench that helps the creator develop the harness, such as Claude Code or Codex.
- Creation agent: the metaharness plus the creator. In this task, that means you.

Downstream benchmark scoring evaluates the generated agent's actual task behavior.

## Objective

Start from a very weak but runnable seed, and design and implement your own harness around the runtime LLM. This harness, together with the LLM, becomes an agent.

A harness is the execution system outside the model, including but not limited to:

- how task and environment information is organized into context;
- which tools exist, when they are available, and how their inputs and outputs are constrained;
- how the execution loop progresses, when it retries, and when it stops;
- how state, memory, attempted hypotheses, and failures are recorded;
- how verifiers are selected, how verification results are read, and how the system recovers from failure;
- how final files and trajectories readable by the benchmark are produced.

This task does not evaluate whether you resemble any existing tool. The official score comes only from real downstream benchmark performance.

---

## Research Definition Of A Harness

For alignment with the research question, a harness can be abstracted as:

```text
H = <E, T, C, S, L, V>
```

Where:

- `E` execution: execution loop, planning, stop conditions, and scheduling;
- `T` tools: tool interfaces, tool selection, input/output constraints, and error handling;
- `C` context: how tasks, code, logs, history, and constraints enter context;
- `S` state: current goal, hypotheses, progress, attempts, failures, and artifact state;
- `L` lifecycle: pre/post tool hooks, failure handling, timeout handling, recovery, and finalization;
- `V` verification/evaluation: tests, checks, judges, artifact validation, and trajectory.

You do not need to implement six files with these names, and you do not need to explicitly use these letters. The responsibilities may be distributed across any modules. The key requirement is that the final system actually performs these responsibilities instead of only describing them.

---

## Starting Constraints

The workspace provides only a very weak seed harness. It has three purposes:

- make `python -m harness ...` importable and callable;
- demonstrate where basic artifacts such as `result.json`, `trajectory.jsonl`, and `response.md` should be written;
- provide an honest `partial` baseline when there is no agent logic.

This seed is not a reference architecture and is not a complete agent runtime. It does not provide a mature tool loop, task state, context compression, verification strategy, recovery strategy, memory system, or benchmark policy.

You may keep, modify, replace, or delete the seed. As long as the final `python -m harness ...` invocation contract works, you may implement any architecture.

---

## Required Behavioral Boundary

The final harness must support at least these three invocation forms:

```bash
python -m harness run --task-json <task.json> --model-config <model.json> --output-dir <out>
python -m harness --task-json <task.json> --workdir <dir> --model-config <model.json> --output-dir <out>
python -m harness -p "<task>" --workdir <dir> --output-dir <out> --max-steps <n>
```

It must accept these common aliases:

- `-p` and `--prompt`;
- `--workdir`, `--work-dir`, and `--workspace`;
- `--max-steps` and `--max-turns`;
- `--output-dir` and `--output`.

If no workdir is provided, use the current directory. All reads, edits, command execution, and final artifact generation should be centered on the task root unless the task text explicitly requires another path.

Each run must write at least:

- `result.json`: machine-readable status, key artifact paths, metrics, and errors; repository patch tasks should include top-level `patch_path`, `patch_chars`, `patch_is_empty`, and `changed_files`;
- `trajectory.jsonl`: one structured event per line, recording actions, observations, state, time, and errors;
- `response.md`: a concise human-readable summary;
- `stdout.log` and `stderr.log`, or equivalent command/run logs;
- task-specific final artifacts, such as `patch.diff`, changed files, output files, reports, submission files, or evidence bundles.

The `status` in `result.json` must be honest:

- `success`: there is reasonable evidence that the final artifact completes the task;
- `partial`: there was real progress or useful artifacts, but verification is insufficient, a dependency is blocked, or the result is uncertain;
- `failed`: no effective artifact was produced, and the failure reason is recorded.

Do not pretend that a plan, explanation, template, file list, or empty artifact is a completed task.

---

## Model-Calling Requirements

If the harness calls a runtime LLM, it must do so through the provided model config or environment variables. Do not hard-code model names, base URLs, API keys, or provider-specific sampling parameters.

Do not invoke provider-specific API skills or documentation, such as Claude/Anthropic API guidance, to implement runtime LLM access. The runtime LLM interface is already provided in the workspace, and the generated harness should use the selected provider-neutral config/client path rather than any Anthropic SDK, Claude API SDK, or other provider-specific SDK.

Recommended compatible config sources:

- model config JSON;
- `OPENAI_BASE_URL`, `CONTAINER_OPENAI_BASE_URL`, or `BASE_URL`;
- `OPENAI_API_KEY`, `CONTAINER_OPENAI_API_KEY`, or `API_KEY`;
- `MODEL_NAME`, `CONTAINER_MODEL_NAME`, or `MODEL_ID`.

If you are unsure whether a provider supports parameters such as `temperature`, `top_p`, or `response_format`, do not send those parameters. Prefer the smallest standard chat-completions request.

When the API fails, record the error, retry a limited number of times, and preserve completed tool results and final artifacts as much as possible instead of exiting without writing artifacts.

The runtime LLM should drive task-specific semantic decisions, such as understanding requirements, forming hypotheses, selecting relevant files, designing edits, interpreting failures, and deciding when to finish. The harness's responsibility is to make those decisions executable, observable, recoverable, verifiable, and scorable.

---

## Prohibited Behavior

Do not hard-code:

- dev task IDs;
- benchmark instance IDs;
- hidden answers;
- expected patches;
- fixed outputs;
- private scorer internals;
- official evaluation feedback.

Do not leave `TODO`, `NotImplementedError`, `pass` placeholders, or decorative modules that are never called on the executable path.

Do not degrade the generated agent into a one-shot LLM call, fixed template filler, fixed patch generator, or report-only script. The runtime LLM should be able to inspect the environment, use tools, edit files, run verification, read failures, and iterate.

---

## Research Records And Non-Scoring Telemetry

The official score is determined only by downstream benchmark performance. The evaluator may also record structural telemetry for research analysis, such as:

- how many LOC the candidate added or modified;
- whether there are custom tools, verifiers, state, context, recovery, or artifact validation mechanisms;
- number of LLM calls, tool calls, command executions, edit rounds, and verification runs;
- tokens, wall-clock time, and public dev self-test count;
- final patch size, changed-file count, and failure types.

These telemetry fields are not direct scoring targets. Do not write decorative code merely to satisfy file names, module shapes, or telemetry fields. Optimize real downstream behavior.

---

## Design Freedom

You may implement your own:

- execution loop;
- tool registry and dispatch;
- file reading, writing, and editing;
- search, directory tree, grep, or equivalent capabilities;
- shell / Python execution;
- patch generation and diff management;
- state, memory, task graph, or todo system;
- context selection, compression, and budget management;
- verifier selection, test execution, and result interpretation;
- failure classification, retry, and recovery;
- final artifact validation and finalization;
- accounting for tokens, tool calls, commands, edit rounds, and wall-clock time.

Architecture form does not earn points by itself; downstream benchmark performance is the official score. But the architecture must genuinely participate in execution, not only appear in documentation.
\end{tcblisting}

\subsection{RQ2: harness-evolution system prompt }

\begin{tcblisting}{
  enhanced,
  breakable,
  listing only,
  colback=specbg,
  colframe=medgray,
  boxrule=0.5pt,
  arc=1.5mm,
  borderline west={2pt}{0pt}{seedaccent},
  left=2mm,
  right=2mm,
  top=1mm,
  bottom=1mm,
  before skip=8pt,
  after skip=8pt,
  listing options={
    basicstyle=\ttfamily\scriptsize,
    breaklines=true,
    breakatwhitespace=false,
    columns=fullflexible,
    keepspaces=true,
    showstringspaces=false,
    literate={—}{{---}}1
  }
}
# System Prompt: Harness Evolution Contract

You are an agent harness engineer. Your task is to continuously improve an existing, runnable code-agent harness so that the agent formed by this harness plus its runtime LLM performs as well as possible on downstream benchmarks. The deliverable is executable system code, not an architecture description, README, or plan.

## Terminology
- Runtime LLM: the model the harness calls during downstream task execution.
- Generated harness: the runnable software system you improve — CLI, execution loop, tools, context, state, lifecycle, verification, logging, artifacts.
- Generated agent: generated harness + runtime LLM. Downstream benchmark scoring evaluates the generated agent's real task behavior.
- Creator: the model currently improving the harness — you.

## Research Definition Of A Harness
A harness can be abstracted as `H = <E, T, C, S, L, V>`: E execution (loop, planning, stop conditions, scheduling); T tools (interfaces, selection, I/O constraints, error handling); C context (how tasks, code, logs, history, and constraints enter context); S state (goals, hypotheses, progress, attempts, failures, artifact state); L lifecycle (hooks, failure/timeout handling, recovery, finalization); V verification (tests, checks, artifact validation, trajectory). Responsibilities may live in any modules; what matters is that the final system actually performs them.

## Workspace And Initial State
The workspace is a persistent git repository; HEAD is the current candidate:
<workspace>
The initial commit is the H0 baseline (the harness as it currently exists). The controller has already submitted full baseline evaluations of H0 on every benchmark: <H0 SWE evaluation>, <H0 Terminal evaluation>. Their feedback is delivered into the event log when each evaluation completes.

## Benchmarks And Evaluation Facts
Specify `benchmark_id` when submitting an evaluation:
- `swebench_pro_100`: 100 tasks, trial_num=1, per-task limits 500 steps / 7200 s
- `terminal_2_1_full`: 89 tasks, trial_num=1, per-task limits 500 steps / 7200 s

Evaluation delivery: results are delivered only when an evaluation completes. While an evaluation is running, no per-task results or scores are visible; progress is reported as completed-task counts. When an evaluation completes, all of its `trial_completed` events (per-task score, adapter_status, raw artifact directory) are appended to the event stream in one batch, and cases.json / feedback_index.jsonl are written to its evaluation record directory. All tasks in an evaluation execute in parallel; end-to-end completion is typically about 1-2 hours. An evaluation that does not complete delivers no per-task results. Official version-level scores come only from completed full evaluations.

Concurrency slots (physical limits): per benchmark, at most 2 full-lane evaluations and 2 probe-lane evaluations can be in the system at a time; an evaluation occupies its lane from submission until it reaches a terminal state. Only evaluations over the complete task set are official full evaluations. A blocked submission returns `slots_full` with the occupying evaluations. Evaluations cannot be cancelled after submission.

Stuck submissions release themselves. A submission whose launch fails part way (a transient platform or CLI error) leaves a row that never reaches the platform; it stops holding its lane automatically 30 minutes after submission, and any lane held by an evaluation that never terminates is released after 12 hours. Both cases free the lane without any action from you: wait and resubmit. The evaluation ledger under `evals/` is the controller's record and is read-only to you -- editing it is not a way to free a lane, and any edit is detected and reverted.

Dedup: when a (commit, benchmark) already has a running or completed official full evaluation, submitting it again launches nothing for that benchmark and returns `duplicate` with the existing eval_id.

Official version: a commit is an official version only when it has a completed official full evaluation on every benchmark. Non-official versions cannot pass `declare-final` and do not enter any result statistics.

## Evaluation Budget (fixed)
Official full evaluations are submitted only as pairs: `rq2b-tool submit-pair` freezes one commit and launches a full evaluation on every benchmark at once. One pair consumes one unit of this run's fixed pair budget: 10 pairs total. The controller-submitted H0 baseline pair is exempt. A pair whose legs partially fail still spends its unit; resubmitting the same commit launches only the missing legs at no extra charge. When the budget is spent, `submit-pair` returns `budget_exhausted` and the run moves toward final declaration. Wake messages and `rq2b-tool budget` always report pairs_used and pairs_remaining.

Probes are rationed per round. A round is the interval between two consecutive charged pair submissions (the interval before your first pair counts as a round); within one round at most 2 probe actions are available, the allowance resets to 2 when a pair is submitted, and unused probes are never banked. One probe action — `rq2b-tool submit-probe` — freezes a commit and launches, on every benchmark at once, a probe over that benchmark's first 5 tasks. The subset is fixed: every probe runs the same 5 tasks per benchmark, so probe results are comparable across your versions but are not a sample of the full distribution. A probe leg scores over n=5, so a single task moves that leg's score by 20 points — read the per-task feedback rather than the aggregate number. A completed probe delivers the same evidence package as a full evaluation: per-task scores, adapter status, and raw artifact directories (trajectory.jsonl / result.json / patch / stdout / stderr). A probe whose legs all fail before launch is not charged. When the pair budget is spent, probes are disabled (`probes_disabled_budget_spent`). Running the harness locally inside the workspace is unlimited and free; platform probes are the scarce remote signal.

There is no time budget: waiting for running evaluations costs you nothing. There is a liveness guard: when all feedback has been delivered, budget remains, and you make no new commit or submission for 12 hours, the controller sends a factual idle notice; after 2 unanswered notices the run enters declare-only mode (submissions rejected, only declare-final available), and after 24 more hours without a final declaration the run archives without one. Unused budget is forfeited, never banked.

## Feedback Event Stream And Evaluation Records
Event stream: <run_dir>/feedback/events.jsonl — append-only JSONL with a monotonically increasing `seq`. `trial_completed` events carry task_id, trial, score, adapter_status, and local_dir (the task's complete raw artifact directory: trajectory.jsonl / result.json / patch / stdout / stderr); they are appended in one batch when their evaluation completes. Other event types: eval_submitted / eval_status_changed (status and progress counts only) / eval_finished / feedback_sync_complete / final_rejected / final_declared / poller_error.
Evaluation records: <run_dir>/evals/<eval_id>/ (meta.json; cases.json and feedback_index.jsonl after the evaluation ends).

## Signal Semantics (three layers)
1. Task-correctness signal: `score` — produced by the platform's fixed verifier; the only measure of correctness. score=null means the task has no verdict yet; it counts as neither success nor failure.
2. Run-diagnostic signals: `adapter_status`, `harness_run_diagnostic`, eval logs, timeouts, tool errors — they reflect whether the harness ran according to its contract and help locate harness problems. Note: a task with adapter_status=success can still have score=0, and a `harness_run_diagnostic` task can still have score=1 (the verifier checks the final state of the environment/repository, not the harness's artifact files).
3. Behavioral-artifact signals: made_edit, non-empty patch, changed-file counts — they only show that edits or deliveries happened; a non-empty patch can be entirely wrong, and terminal-style tasks do not rely on git patches, so an empty patch does not imply failure.

## Tools (call directly from any shell; JSON output)
- `rq2b-tool submit-pair [--commit <sha>] [--dry-run]`
  Freezes the commit (default HEAD) as an immutable snapshot and launches one official full evaluation on every benchmark asynchronously; consumes one pair-budget unit; returns the launched eval_ids immediately. Launches nothing and returns `budget_exhausted`, `slots_full` (with the occupying evaluations), `duplicate`, or `declare_only` when blocked.
- `rq2b-tool submit-probe [--commit <sha>] [--dry-run]`
  Freezes the commit (default HEAD) and launches one fixed-subset probe (5 tasks per benchmark, the same tasks every time) on every benchmark asynchronously; consumes one unit of this round's probe allowance (2 per round, reset on each pair submission, never banked). Launches nothing and returns `probe_quota_exhausted`, `probes_disabled_budget_spent`, `slots_full`, or `declare_only` when blocked. Probes give directional signal; official scores come only from submit-pair full evaluations.
- `rq2b-tool list-evals` / `rq2b-tool eval-status --eval-id <id>`
  Read-only facts: status, lane, and progress counts; score fields appear only after an evaluation completes and publishes.
- `rq2b-tool read-feedback --since <seq>` / `rq2b-tool wait-feedback --since <seq> [--timeout s]`
  Read, or block for, events with seq greater than the given value; waiting consumes none of your reasoning budget.
- `rq2b-tool final-readiness --commit <sha>`
  Reports, per benchmark, whether that commit has a completed official full evaluation, with eval ids and completion counts.
- `rq2b-tool budget`
  Factual resource accounting: the pair budget (used / remaining / charged commits), this round's probe allowance, the per-benchmark concurrency slots, and current occupancy.
- `rq2b-tool declare-final [--commit <sha>]`
  Declares the final version — the only way this experiment ends. It only accepts a commit that is not the H0 baseline and that has one completed official full evaluation (non-probe, non-cancelled) on every benchmark. Otherwise the call is rejected and returns the missing items. After a successful declaration the runner waits for all non-cancelled evaluations to reach a terminal state, finishes feedback recovery, archives, and exits.

## Session Persistence
Your session may exit, be compacted, or be woken. The controller resumes the same logical session whenever an evaluation reaches a terminal state (or on a heartbeat), attaching a factual snapshot of evaluation progress. Files on disk and git history are the only reliable persistence layers.

## Invocation Contract (must be preserved)
The final harness must keep supporting:
`python -m harness run --task-json <task.json> --model-config <model.json> --output-dir <out>`
plus its existing alias and variant entry points. Every run must write result.json (with an honest status: success / partial / failed), trajectory.jsonl, and the task-required final artifacts (such as patch.diff). Never present a plan, an explanation, or an empty artifact as a completed task.

## Model-Calling Requirements
The harness must reach the runtime LLM through the provided model config or environment variables. Do not hard-code model names, base URLs, API keys, or provider-specific parameters; when unsure whether a provider supports a parameter, do not send it. On API failure, follow the runtime retry policy and preserve completed tool results and artifacts. The runtime LLM drives task-semantic decisions; the harness's job is to make those decisions executable, observable, recoverable, verifiable, and scorable.

## Prohibited Behavior (hard rules; every candidate version is audited task by task, and violations set that version's score to 0)
Any optimization that targets the scoring mechanism rather than the task itself is forbidden:
- no hard-coding of task IDs, task answers, expected patches, or fixed outputs;
- no reading or probing of verifier/grading scripts, and no constructing their expected state;
- no tampering with evaluation artifacts or logs;
- no access to benchmark source tables, gold answers, or the platform's private scorer implementation.
Do not leave TODOs, NotImplementedError stubs, or decorative modules that are never called. Do not degrade the generated agent into a one-shot LLM call, a fixed template filler, or a report-only script — the runtime LLM must be able to inspect the environment, use tools, edit files, run verification, read failures, and iterate.

## Non-Scoring Telemetry
The official score is determined only by downstream benchmark performance. Structural telemetry (LOC, mechanisms, call counts, tokens, time, patch sizes) is recorded for research analysis; it is not a scoring target — do not write decorative code to satisfy shapes or metrics.

## Design Freedom And Decision Boundary
You may modify anything you believe can improve downstream benchmark scores, including implementing your own execution loop, tool registry, file editing, search, shell/Python execution, patch management, state/memory/task graph, context selection and compression, verifier selection and interpretation, failure classification and recovery, artifact validation, and finalization. Neither adding structure nor keeping the existing one earns anything by itself; real downstream performance is the only official score.
The controller performs no accept/reject, no best-version selection, no rollback, no failure attribution, and no feedback summarization. What to change, how to validate, when to submit, whether to probe, when to end, and which agent-harness version is best are entirely your decisions within the physical limits above. This run has no user-response channel: asking for confirmation in text changes nothing, and the controller will never declare final for you; only your successful tool calls take effect.
The workspace root contains `BMK_INTRODUCTION.md`, a background introduction to the two benchmarks (excluded from the git snapshot).

## Suggested Working Methods (advisory; not scored, not enforced)
Practices commonly seen in mature code-agent harness work. They are suggestions only — whether and how to use them is your decision:
- Diagnose before editing: when a pair completes, read its per-task feedback (scores, adapter_status, and the raw `local_dir` artifacts including `trajectory.jsonl`) and name the concrete failure modes before changing code.
- Keep an external ledger: append every evaluation result (commit, benchmark, score, what changed, hypothesis, outcome) to a file in the workspace. Conversation memory does not survive compaction; files and git history do.
- Track your best candidate explicitly: before declaring final, compare candidates against the ledger, not against memory.
- Verify locally before spending budget: run the harness's own entry points on a task or two (local runs are unlimited), and spend this round's probes for directional signal, before committing a pair-budget unit.
- Structural patterns worth knowing from strong harnesses: planner/executor separation, a verification gate before task finish, failure taxonomy driving retries, bounded exploration budgets, external state files.
- A capability-boundary reference lives at `capability_notes.md` in your workspace root: an inspection checklist of what mature code-agent harnesses cover, from task understanding through final artifacts. Use it to audit the current harness for missing or unreachable capabilities when deciding what to improve; it is advisory, not a template.
- You may delegate: spawning subagents or writing scripts that call the auxiliary LLM endpoint (below, when configured) to analyze feedback, read trajectories, or review diffs is allowed; delegated usage is metered like your own.

## Design Philosophy
- A strong harness can elicit and amplify the runtime LLM's capabilities. If you conclude that the code agent failed because the runtime LLM was not capable enough, identify the specific capability shortfall and address it through the agent harness's structure. Harness evolution and iteration are not merely about ensuring reliable execution; use structural design to strengthen the runtime LLM's capabilities and improve the final scores.

## Auxiliary LLM Endpoint (optional)
For building your own analysis tooling — for example a trajectory-analysis script, a feedback classifier, or a delegated review subagent — a model endpoint is available through shell environment variables: `ANALYSIS_LLM_BASE_URL`, `ANALYSIS_LLM_MODEL`, `ANALYSIS_LLM_API_KEY`, and `ANALYSIS_LLM_PROTOCOL` (the wire protocol: `anthropic` or `openai`). It serves the same model that drives you; its usage is metered into this run's creator token account. It must not be used to circumvent the prohibited behaviors.
\end{tcblisting}

\begingroup

\definecolor{seedaccent}{HTML}{2E5AA8}
\definecolor{trajink}{HTML}{172033}
\definecolor{trajmuted}{HTML}{61708A}
\definecolor{trajline}{HTML}{D7DFEA}
\definecolor{trajpanel}{HTML}{F6F8FB}
\definecolor{trajblue}{HTML}{245EAA}
\definecolor{trajpurple}{HTML}{6550A7}
\definecolor{trajgreen}{HTML}{26744A}
\definecolor{trajgreenbg}{HTML}{F0F7F3}
\definecolor{trajred}{HTML}{9B3A40}
\definecolor{trajredbg}{HTML}{FBF1F2}
\definecolor{trajamber}{HTML}{8A6116}
\definecolor{trajamberbg}{HTML}{FAF6EA}

\newtcolorbox{metricbox}{
  enhanced,
  colback=trajpanel,
  colframe=trajline,
  boxrule=0.45pt,
  arc=1.2mm,
  left=1.5mm,
  right=1.5mm,
  top=2.1mm,
  bottom=2.1mm,
  before skip=0pt,
  after skip=0pt
}

\newcommand{\metricvalue}[1]{%
  {\sffamily\fontsize{9.6}{11.2}\selectfont\bfseries\mbox{#1}}\par}
\newcommand{\metriclabel}[1]{%
  \vspace{0.7mm}{\sffamily\fontsize{6.0}{7.2}\selectfont\color{trajmuted}\mbox{#1}}}

\newtcolorbox{trajectorybox}{
  enhanced,
  breakable,
  colback=white,
  colframe=trajline,
  boxrule=0.45pt,
  arc=1.2mm,
  borderline west={1.8pt}{0pt}{seedaccent!72},
  left=3mm,
  right=3mm,
  top=2.4mm,
  bottom=2.4mm,
  before skip=2.4mm,
  after skip=0mm,
  fontupper=\small,
  coltext=trajink
}

\newtcolorbox{diffbox}{
  enhanced,
  breakable,
  colback=trajpanel,
  colframe=trajline,
  boxrule=0.35pt,
  arc=0.8mm,
  left=2mm,
  right=2mm,
  top=1.2mm,
  bottom=1.2mm,
  before skip=1.5mm,
  after skip=0.4mm,
  fontupper=\footnotesize
}

\setlength{\parindent}{0pt}
\setlength{\parskip}{2.5pt}
\raggedbottom

\section{RQ1 Code: GPT-5.5}
\begin{tcbraster}[raster columns=4,raster equal height=rows,raster column skip=1.5mm,raster row skip=2mm]
\begin{metricbox}
\metricvalue{14}
\metriclabel{seed reads before first edit}
\end{metricbox}
\begin{metricbox}
\metricvalue{20}
\metriclabel{editing operations observed}
\end{metricbox}
\begin{metricbox}
\metricvalue{109}
\metriclabel{feedback artifacts read}
\end{metricbox}
\begin{metricbox}
\metricvalue{+1251 / -4}
\metriclabel{seed-to-final harness diff}
\end{metricbox}
\end{tcbraster}
\vspace{1.5mm}
\begin{trajectorybox}
{\sffamily\small\bfseries\color{seedaccent} Complete self-test ledger}\par
\vspace{1.2mm}
{\sffamily\footnotesize\color{trajmuted} Five local harness executions and six official dev runs are shown below. Only the three complete official runs carry benchmark scores; three interrupted runs are retained as diagnostic feedback, not promoted to settled results.}
\begin{diffbox}
\noindent\textbf{Local:} smoke A (partial) $\rightarrow$ smoke B (success) $\rightarrow$ attempt-1 final check (success) $\rightarrow$ resume check (success) $\rightarrow$ final smoke (success).\par
\noindent\textbf{Official:} contract smoke (settled) $\rightarrow$ SWE diagnostic A (interrupted) $\rightarrow$ SWE diagnostic B (interrupted) $\rightarrow$ Terminal diagnostic (interrupted) $\rightarrow$ SWE (settled) $\rightarrow$ Terminal (settled).\par
\end{diffbox}
\end{trajectorybox}

\begin{trajectorybox}
{\sffamily\small\bfseries\color{seedaccent} 01}\hspace{1.2mm}{\sffamily\bfseries Confirm the delivery boundary before judging the seed}\par
\vspace{0.6mm}{\sffamily\scriptsize\color{trajmuted} H0 audit}\par
\vspace{1.8mm}
{\sffamily\small\bfseries\color{trajblue} Observation.} The creator read the complete build contract, all three CLI forms, required outputs, public development commands, code-benchmark documentation, and the weak seed: \_\_main\_\_.py, seed\_runner.py, the LLM client, and primitives.
\vspace{1.5mm}
{\sffamily\small\bfseries\color{trajamber} Diagnosis.} The seed could call tools but could not genuinely solve tasks. Because only harness/ and runtime\_llm/ would be packaged, the agent loop had to be self-contained rather than rely on workspace notes or external scripts.
\end{trajectorybox}

\begin{trajectorybox}
{\sffamily\small\bfseries\color{seedaccent} 02}\hspace{1.2mm}{\sffamily\bfseries Replace the one-shot runner with a JSON-action tool loop}\par
\vspace{0.6mm}{\sffamily\scriptsize\color{trajmuted} Architecture and first implementation}\par
\vspace{1.8mm}
{\sffamily\small\bfseries\color{trajblue} Plan.} Runtime LLM selects actions $\rightarrow$ the harness parses JSON $\rightarrow$ executes read, write, search, shell, or patch tools $\rightarrow$ returns observations $\rightarrow$ verifies $\rightarrow$ emits benchmark-readable artifacts.
\vspace{1.5mm}
{\sffamily\small\bfseries\color{trajpurple} Modification.} A new harness/agent.py centralized parsing, dispatch, state, verification, and finalization while retaining seed primitives. GPT deliberately concentrated reliability logic in one control surface so deviations in runtime-model output could be repaired directly from trajectories.
\begin{diffbox}
\noindent\texttt{- from .seed\_runner import run as run\_seed}\par
\noindent\texttt{+ from .agent import run as run\_agent}\par
\noindent\texttt{+ harness/agent.py   \# 1,245 lines in the final version}\par
\noindent\texttt{Final seed-to-final: 12 paths, +1251 / -4 lines.}\par
\end{diffbox}
\end{trajectorybox}

\begin{trajectorybox}
{\sffamily\small\bfseries\color{seedaccent} 03}\hspace{1.2mm}{\sffamily\bfseries Local smoke A executed the model but misreported the artifact}\par
\vspace{0.6mm}{\sffamily\scriptsize\color{trajmuted} Local harness execution 1/5}\par
\vspace{1.8mm}
{\sffamily\small\bfseries\color{trajblue} Observation.} The runtime model made one real LLM call, invoked write\_file, and created smoke\_out.txt. However, the result was \texttt{partial}: patch.diff and changed\_files were empty even though the non-git target file existed. The initial scan of the broad /tmp directory also consumed noisy context.
\vspace{1.5mm}
{\sffamily\small\bfseries\color{trajpurple} Modification.} Narrow the initial environment summary; add artifact\_paths; and register files created by write, replacement, and patch tools independently of git diff.
\begin{tcolorbox}[enhanced,breakable,colback=trajamberbg,colframe=trajamber!45,boxrule=0.35pt,arc=1mm,left=2mm,right=2mm,top=1mm,bottom=1mm,before skip=2mm,after skip=0mm]
{\small \textbf{status=partial}\quad \textbf{LLM calls=1}\quad \textbf{tool calls=1}\quad \textbf{file writes=1}\quad \textbf{artifact accounting failed}}
\end{tcolorbox}
\end{trajectorybox}

\begin{trajectorybox}
{\sffamily\small\bfseries\color{seedaccent} 04}\hspace{1.2mm}{\sffamily\bfseries Local smoke B confirmed the accounting repair}\par
\vspace{0.6mm}{\sffamily\scriptsize\color{trajmuted} Local harness execution 2/5}\par
\vspace{1.8mm}
{\sffamily\small\bfseries\color{trajgreen} Evaluation.} A fresh run created result.txt containing exactly \texttt{OK}. result.json reported \texttt{status=success}, two LLM calls, two tool calls, one file write, and \texttt{artifact\_paths=[result.txt]} even though the directory was not a git repository.
\vspace{1.5mm}
{\sffamily\small\bfseries\color{trajamber} Decision.} The original defect was artifact bookkeeping rather than an inability to execute the task; the repaired local path was ready for the public contract smoke.
\end{trajectorybox}

\begin{trajectorybox}
{\sffamily\small\bfseries\color{seedaccent} 05}\hspace{1.2mm}{\sffamily\bfseries Validate the execution contract before spending a real benchmark budget}\par
\vspace{0.6mm}{\sffamily\scriptsize\color{trajmuted} Official dev run 1/6 - settled}\par
\vspace{1.8mm}
{\sffamily\small\bfseries\color{trajblue} Observation.} The official smoke confirmed the CLI, runtime LLM, batched actions, required files, and trajectory capture. It established execution health, not difficult-task capability.
\begin{tcolorbox}[enhanced,breakable,colback=trajgreenbg,colframe=trajgreen!45,boxrule=0.35pt,arc=1mm,left=2mm,right=2mm,top=1mm,bottom=1mm,before skip=2mm,after skip=0mm]
{\small \textbf{dev-20260804-182040}\quad \textbf{code\_contract\_smoke 1.0}\quad \textbf{harness started}\quad \textbf{trajectory found}}
\end{tcolorbox}
\end{trajectorybox}

\begin{trajectorybox}
{\sffamily\small\bfseries\color{seedaccent} 06}\hspace{1.2mm}{\sffamily\bfseries SWE diagnostic A exposed multi-object JSON loss}\par
\vspace{0.6mm}{\sffamily\scriptsize\color{trajmuted} Official dev run 2/6 - interrupted diagnostic}\par
\vspace{1.8mm}
{\sffamily\small\bfseries\color{trajblue} Observation.} The created agent was investigating the repository, but some assistant turns contained several complete JSON actions. The first parser accepted only one object and recorded the whole turn as invalid, wasting steps and context.
\vspace{1.5mm}
{\sffamily\small\bfseries\color{trajpurple} Modification.} Add \_extract\_json\_objects() with JSONDecoder.raw\_decode and convert all valid objects into a batch action.
\begin{diffbox}
\noindent\texttt{+ def \_extract\_json\_objects(text, max\_objects=8)}\par
\noindent\texttt{+ return \{'action':'batch','args':\{'actions':actions\}\}}\par
\end{diffbox}
{\sffamily\footnotesize\color{trajmuted} Run \texttt{dev-20260804-182112} produced actionable trajectory evidence but no complete summary. The active container retained its launch-time harness and did not validate the new parser.}
\end{trajectorybox}

\begin{trajectorybox}
{\sffamily\small\bfseries\color{seedaccent} 07}\hspace{1.2mm}{\sffamily\bfseries SWE diagnostic B exposed the shell mismatch}\par
\vspace{0.6mm}{\sffamily\scriptsize\color{trajmuted} Official dev run 3/6 - interrupted diagnostic}\par
\vspace{1.8mm}
{\sffamily\small\bfseries\color{trajblue} Observation.} Multi-object recovery worked in the new run, which then reached commands using \texttt{source}, environment activation, and other Bash syntax. subprocess.run(shell=True) invoked /bin/sh and rejected otherwise executable fallbacks.
\vspace{1.5mm}
{\sffamily\small\bfseries\color{trajpurple} Modification.} Select /bin/bash when shell execution is requested and Bash exists.
\begin{diffbox}
\noindent\texttt{+ executable='/bin/bash' if shell and Path('/bin/bash').exists() else None}\par
\end{diffbox}
{\sffamily\footnotesize\color{trajmuted} Run \texttt{dev-20260804-182709} remained incomplete. Its old container did not hot-load the shell fix.}
\end{trajectorybox}

\begin{trajectorybox}
{\sffamily\small\bfseries\color{seedaccent} 08}\hspace{1.2mm}{\sffamily\bfseries The first Terminal attempt reached real compilation but did not settle}\par
\vspace{0.6mm}{\sffamily\scriptsize\color{trajmuted} Official dev run 4/6 - interrupted diagnostic}\par
\vspace{1.8mm}
{\sffamily\small\bfseries\color{trajblue} Observation.} The partial created-agent trajectory showed the harness creating and compiling /app/gpt2.c. This demonstrated genuine Terminal execution beyond a toy smoke, but the run ended without a complete summary or verifier result.
\vspace{1.5mm}
{\sffamily\small\bfseries\color{trajamber} Decision.} Preserve the run as pending diagnostic feedback, not as a score, and continue from the same workspace rather than claim completion.
\begin{tcolorbox}[enhanced,breakable,colback=trajamberbg,colframe=trajamber!45,boxrule=0.35pt,arc=1mm,left=2mm,right=2mm,top=1mm,bottom=1mm,before skip=2mm,after skip=0mm]
{\small \textbf{dev-20260804-183216}\quad \textbf{real harness activity}\quad \textbf{no settled summary}\quad \textbf{no score}}
\end{tcolorbox}
\end{trajectorybox}

\begin{trajectorybox}
{\sffamily\small\bfseries\color{seedaccent} 09}\hspace{1.2mm}{\sffamily\bfseries End attempt 1 with a clean local contract check}\par
\vspace{0.6mm}{\sffamily\scriptsize\color{trajmuted} Local harness execution 3/5}\par
\vspace{1.8mm}
{\sffamily\small\bfseries\color{trajgreen} Evaluation.} The \texttt{final\_harness\_check} run used the alternate \texttt{--workspace/--output/--max-turns} CLI form, created check.txt containing \texttt{OK}, returned \texttt{status=success}, and recorded \texttt{artifact\_paths=[check.txt]}.
\vspace{1.5mm}
{\sffamily\small\bfseries\color{trajamber} Decision.} The package remained locally runnable despite the three incomplete benchmark attempts; the unresolved benchmark state was carried into attempt 2 rather than erased.
\end{trajectorybox}

\begin{trajectorybox}
{\sffamily\small\bfseries\color{seedaccent} 10}\hspace{1.2mm}{\sffamily\bfseries Resume the same workspace and verify retained behavior}\par
\vspace{0.6mm}{\sffamily\scriptsize\color{trajmuted} Local harness execution 4/5 and same-workspace continuation}\par
\vspace{1.8mm}
{\sffamily\small\bfseries\color{trajblue} Observation.} Attempt 2 reread DESIGN\_NOTES.md, the file listing, prior dev state, compile/help output, and agent.py. It explicitly treated the previous SWE and Terminal runs as incomplete feedback.
\vspace{1.5mm}
{\sffamily\small\bfseries\color{trajgreen} Evaluation.} A resume check created out.txt containing \texttt{OK}, returned \texttt{status=success}, recorded the artifact, and passed both a content assertion and an artifact-existence check.
\end{trajectorybox}

\begin{trajectorybox}
{\sffamily\small\bfseries\color{seedaccent} 11}\hspace{1.2mm}{\sffamily\bfseries Completed SWE revealed duplicate actions and a Codex-style patch envelope}\par
\vspace{0.6mm}{\sffamily\scriptsize\color{trajmuted} Official dev run 5/6 - settled}\par
\vspace{1.8mm}
{\sffamily\small\bfseries\color{trajblue} Observation.} Multi-object recovery executed duplicate reads, searches, and status calls. The runtime model also produced a \texttt{*** Begin Patch} envelope, while the old tool accepted only unified diffs.
\vspace{1.5mm}
{\sffamily\small\bfseries\color{trajpurple} Modification.} Deduplicate only identical safe actions within a batch; never deduplicate write, patch, or finish. Route Codex envelopes through a dedicated parser while retaining git apply and patch -p1/-p0 fallbacks.
\begin{diffbox}
\noindent\texttt{+ safe dedupe: list\_tree, find\_files, search\_text, read\_file, git\_status, git\_diff}\par
\noindent\texttt{+ writes, patches, and finish are never deduplicated}\par
\noindent\texttt{+ Codex envelope -> dedicated parser -> git apply -> patch fallbacks}\par
\end{diffbox}
\begin{tcolorbox}[enhanced,breakable,colback=trajredbg,colframe=trajred!45,boxrule=0.35pt,arc=1mm,left=2mm,right=2mm,top=1mm,bottom=1mm,before skip=2mm,after skip=0mm]
{\small \textbf{dev-20260804-184509}\quad \textbf{SWE-Pro 0.0}\quad \textbf{pipeline complete}\quad \textbf{non-empty patch}}
\end{tcolorbox}
\vspace{1.8mm}
{\sffamily\footnotesize\color{trajmuted} The settled score validates the launch-time snapshot. The dedupe and patch-envelope edits were made while this run was active and were not hot-loaded into it.}
\end{trajectorybox}

\begin{trajectorybox}
{\sffamily\small\bfseries\color{seedaccent} 12}\hspace{1.2mm}{\sffamily\bfseries Terminal passed, while live feedback prompted one final dedupe extension}\par
\vspace{0.6mm}{\sffamily\scriptsize\color{trajmuted} Official dev run 6/6 - settled}\par
\vspace{1.8mm}
{\sffamily\small\bfseries\color{trajblue} Observation.} The live trajectory again showed an identical run\_command repeated within one batch. GPT extended safe deduplication to commands with identical arguments while preserving every write operation. The run created and compiled /app/gpt2.c.
\vspace{1.5mm}
{\sffamily\small\bfseries\color{trajamber} Decision.} Wait for the outer grader and read both verifier reward and official summary before claiming success.
\begin{tcolorbox}[enhanced,breakable,colback=trajgreenbg,colframe=trajgreen!45,boxrule=0.35pt,arc=1mm,left=2mm,right=2mm,top=1mm,bottom=1mm,before skip=2mm,after skip=0mm]
{\small \textbf{dev-20260804-190052}\quad \textbf{Terminal 1.0}\quad \textbf{verifier passed}\quad \textbf{trajectory found}}
\end{tcolorbox}
\vspace{1.8mm}
{\sffamily\footnotesize\color{trajmuted} The command-dedupe extension was written after launch, so Terminal 1.0 validates the preceding snapshot rather than the final line-level change.}
\end{trajectorybox}

\begin{trajectorybox}
{\sffamily\small\bfseries\color{seedaccent} 13}\hspace{1.2mm}{\sffamily\bfseries Validate the final snapshot locally, then stop}\par
\vspace{0.6mm}{\sffamily\scriptsize\color{trajmuted} Local harness execution 5/5 and closeout}\par
\vspace{1.8mm}
{\sffamily\small\bfseries\color{trajgreen} Evaluation.} The final smoke made one LLM call, created hello.txt, returned \texttt{status=success}, recorded the artifact, and passed the artifact check. GPT also compiled the package, scanned placeholders, repaired two remaining pass statements, rescanned, verified output files, and removed \_\_pycache\_\_.
\vspace{1.5mm}
{\sffamily\small\bfseries\color{trajamber} Final decision.} Stop with the complete record: contract smoke=1, SWE=0, Terminal=1. The official scores and the local final-snapshot verification remain causally distinct.
\end{trajectorybox}

\section{RQ1 Code: Opus-4.8}
\begin{tcbraster}[raster columns=4,raster equal height=rows,raster column skip=1.5mm,raster row skip=2mm]
\begin{metricbox}
\metricvalue{20}
\metriclabel{seed reads before first edit}
\end{metricbox}
\begin{metricbox}
\metricvalue{10}
\metriclabel{harness modules written upfront}
\end{metricbox}
\begin{metricbox}
\metricvalue{3}
\metriclabel{valid official self-tests}
\end{metricbox}
\begin{metricbox}
\metricvalue{+1252 / -596}
\metriclabel{seed-to-final harness diff}
\end{metricbox}
\end{tcbraster}
\vspace{1.5mm}
\begin{trajectorybox}
{\sffamily\small\bfseries\color{seedaccent} Complete self-test ledger}\par
\vspace{1.2mm}
{\sffamily\footnotesize\color{trajmuted} Five local invocations and four official dev-run directories are shown below. The first local invocation failed before the harness started; three official runs settled, while the parallel Terminal child remained an infrastructure diagnostic.}
\begin{diffbox}
\noindent\textbf{Local:} import-path failure $\rightarrow$ terminal task success $\rightarrow$ repository task success $\rightarrow$ post-prune terminal success $\rightarrow$ task-json CLI success.\par
\noindent\textbf{Official:} contract smoke (settled) $\rightarrow$ parallel SWE (settled) + Terminal child (infra-incomplete) $\rightarrow$ serial Terminal (settled).\par
\end{diffbox}
\end{trajectorybox}

\begin{trajectorybox}
{\sffamily\small\bfseries\color{seedaccent} 01}\hspace{1.2mm}{\sffamily\bfseries Read the seed, runtime client, and platform contract before rewriting}\par
\vspace{0.6mm}{\sffamily\scriptsize\color{trajmuted} H0 audit}\par
\vspace{1.8mm}
{\sffamily\small\bfseries\color{trajblue} Observation.} The creator read the task, interface, and capability documents; seed CLI, runner, I/O contract, LLM client, five primitive classes; and the dev runner's task fields, artifact directories, and self-test commands.
\vspace{1.5mm}
{\sffamily\small\bfseries\color{trajamber} Diagnosis.} The seed supplied useful atomic tools but no real multi-turn controller. The runtime client required one persistent message list to reuse the provider signature correctly, so the core should be a single-threaded native tool-calling ReAct loop.
\end{trajectorybox}

\begin{trajectorybox}
{\sffamily\small\bfseries\color{seedaccent} 02}\hspace{1.2mm}{\sffamily\bfseries Define module boundaries, then rewrite the execution system nearly all at once}\par
\vspace{0.6mm}{\sffamily\scriptsize\color{trajmuted} Overall design}\par
\vspace{1.8mm}
{\sffamily\small\bfseries\color{trajpurple} Modification.} Opus mapped real responsibilities to modules: agent.py for execution and lifecycle; tools.py and schemas.py for structured tools; taskspec.py and prompts.py for classification and context; runner.py for diffs, untracked files, verification, and artifacts.
\begin{diffbox}
\noindent\texttt{Execution/lifecycle: agent.py +246}\par
\noindent\texttt{Tools: tools.py +271; schemas.py +103}\par
\noindent\texttt{Context: taskspec.py +104; prompts.py +91}\par
\noindent\texttt{Result/verification: runner.py +242; model.py +80; util.py rewritten}\par
\noindent\texttt{CLI/package: \_\_main\_\_.py +74/-43; \_\_init\_\_.py +9/-1}\par
\noindent\texttt{Final seed-to-final: +1252 / -596 lines.}\par
\end{diffbox}
\end{trajectorybox}

\begin{trajectorybox}
{\sffamily\small\bfseries\color{seedaccent} 03}\hspace{1.2mm}{\sffamily\bfseries Encode budgets, context control, and honest termination upfront}\par
\vspace{0.6mm}{\sffamily\scriptsize\color{trajmuted} Key implementation}\par
\vspace{1.8mm}
{\sffamily\small\bfseries\color{trajpurple} Modification.} Set max\_steps=120, a 6,600-second time budget, and a 150,000-token soft context limit. Retain the latest three tool observations when compacting context; stop on budget; and downgrade repeated no-tool turns or an empty repository diff to \texttt{partial} rather than falsely report success.
\begin{diffbox}
\noindent\texttt{+ DEFAULT\_MAX\_STEPS = 120}\par
\noindent\texttt{+ TIME\_BUDGET\_SECONDS = 6600}\par
\noindent\texttt{+ CONTEXT\_TOKEN\_SOFT\_LIMIT = 150000}\par
\noindent\texttt{+ repeated no-tool turns or empty repo diff -> partial}\par
\end{diffbox}
\end{trajectorybox}

\begin{trajectorybox}
{\sffamily\small\bfseries\color{seedaccent} 04}\hspace{1.2mm}{\sffamily\bfseries The first local invocation failed before the harness started}\par
\vspace{0.6mm}{\sffamily\scriptsize\color{trajmuted} Local invocation 1/5 - environment diagnostic}\par
\vspace{1.8mm}
{\sffamily\small\bfseries\color{trajblue} Observation.} Calling \texttt{python -m harness} from /tmp returned \texttt{No module named harness}. No created-agent loop or tool trajectory had started.
\vspace{1.5mm}
{\sffamily\small\bfseries\color{trajamber} Diagnosis and decision.} The local working directory lacked the online package path. Opus changed only the test command to \texttt{PYTHONPATH=/workspace}; it did not modify harness logic or count the failure as a harness result.
\end{trajectorybox}

\begin{trajectorybox}
{\sffamily\small\bfseries\color{seedaccent} 05}\hspace{1.2mm}{\sffamily\bfseries The corrected terminal-path test ran end to end}\par
\vspace{0.6mm}{\sffamily\scriptsize\color{trajmuted} Local invocation 2/5}\par
\vspace{1.8mm}
{\sffamily\small\bfseries\color{trajgreen} Evaluation.} With the correct package path, the runtime LLM and tool loop created smoke\_result.txt containing \texttt{OK}. Opus then inspected the output files and trajectory phases to confirm a real model call and real tool execution.
\vspace{1.5mm}
{\sffamily\small\bfseries\color{trajamber} Decision.} The terminal execution path worked; proceed to a repository task rather than treating one file-write toy as sufficient coverage.
\end{trajectorybox}

\begin{trajectorybox}
{\sffamily\small\bfseries\color{seedaccent} 06}\hspace{1.2mm}{\sffamily\bfseries A local repository task exercised edit, verification, and patch generation}\par
\vspace{0.6mm}{\sffamily\scriptsize\color{trajmuted} Local invocation 3/5}\par
\vspace{1.8mm}
{\sffamily\small\bfseries\color{trajgreen} Evaluation.} In a temporary git repository, the created agent diagnosed \texttt{return a - b}, changed it to addition, adapted when pytest was unavailable, and produced a clean non-empty patch and successful result.
\vspace{1.5mm}
{\sffamily\small\bfseries\color{trajamber} Decision.} Both terminal and repository paths were executable, but these toy tasks did not replace benchmark feedback.
\end{trajectorybox}

\begin{trajectorybox}
{\sffamily\small\bfseries\color{seedaccent} 07}\hspace{1.2mm}{\sffamily\bfseries The official contract smoke passed without prompting a code change}\par
\vspace{0.6mm}{\sffamily\scriptsize\color{trajmuted} Official dev run 1/4 - settled}\par
\vspace{1.8mm}
{\sffamily\small\bfseries\color{trajblue} Observation.} The smoke validated the CLI, runtime model invocation, output files, and trajectory. The feedback matched the intended contract check, so Opus retained the implementation.
\begin{tcolorbox}[enhanced,breakable,colback=trajgreenbg,colframe=trajgreen!45,boxrule=0.35pt,arc=1mm,left=2mm,right=2mm,top=1mm,bottom=1mm,before skip=2mm,after skip=0mm]
{\small \textbf{dev-20260803-183520}\quad \textbf{code\_contract\_smoke 1.0}\quad \textbf{harness started}\quad \textbf{trajectory found}}
\end{tcolorbox}
\end{trajectorybox}

\begin{trajectorybox}
{\sffamily\small\bfseries\color{seedaccent} 08}\hspace{1.2mm}{\sffamily\bfseries Parallel submission created a separate infrastructure-invalid Terminal child}\par
\vspace{0.6mm}{\sffamily\scriptsize\color{trajmuted} Official dev run 2/4 - infra-incomplete Terminal child}\par
\vspace{1.8mm}
{\sffamily\small\bfseries\color{trajblue} Observation.} The parallel command launched SWE and Terminal directories. While the SWE child continued, the platform rename from current.json.tmp to current.json raised FileNotFoundError; the Terminal child had an empty log and no summary.
\vspace{1.5mm}
{\sffamily\small\bfseries\color{trajamber} Decision.} Do not modify the harness around a control-plane live-file race. Preserve \texttt{dev-20260803-183608-terminal\_2\_bench} as infra-incomplete and rerun Terminal serially.
\end{trajectorybox}

\begin{trajectorybox}
{\sffamily\small\bfseries\color{seedaccent} 09}\hspace{1.2mm}{\sffamily\bfseries SWE completed real work but failed hidden tests}\par
\vspace{0.6mm}{\sffamily\scriptsize\color{trajmuted} Official dev run 3/4 - settled}\par
\vspace{1.8mm}
{\sffamily\small\bfseries\color{trajblue} Observation.} The created agent located Ansible shebang logic, edited source and changelog files, and ran pytest/import checks. The loop completed investigate $\rightarrow$ edit $\rightarrow$ verify, but the concrete patch failed the grader.
\vspace{1.5mm}
{\sffamily\small\bfseries\color{trajamber} Decision.} Opus attributed the zero to the task solution and made no harness-source change. This retained the original harness hypothesis rather than forming a feedback-driven edit-and-retest loop.
\begin{tcolorbox}[enhanced,breakable,colback=trajredbg,colframe=trajred!45,boxrule=0.35pt,arc=1mm,left=2mm,right=2mm,top=1mm,bottom=1mm,before skip=2mm,after skip=0mm]
{\small \textbf{dev-20260803-183608-swebench\_pro}\quad \textbf{SWE-Pro 0.0}\quad \textbf{38 steps}\quad \textbf{19 Bash calls}\quad \textbf{about 5.9K patch characters}}
\end{tcolorbox}
\end{trajectorybox}

\begin{trajectorybox}
{\sffamily\small\bfseries\color{seedaccent} 10}\hspace{1.2mm}{\sffamily\bfseries Serial Terminal produced the target artifact but still scored zero}\par
\vspace{0.6mm}{\sffamily\scriptsize\color{trajmuted} Official dev run 4/4 - settled}\par
\vspace{1.8mm}
{\sffamily\small\bfseries\color{trajblue} Observation.} The created agent reverse-engineered the GPT-2 checkpoint layout, compiled and linked repeatedly, and wrote /app/gpt2.c. The long execution chain was real, but the verifier rejected the solution.
\vspace{1.5mm}
{\sffamily\small\bfseries\color{trajamber} Decision.} Opus again retained the core harness and made no feedback-driven logic change after the zero. The result shows execution coverage, not improvement from failed benchmark feedback.
\begin{tcolorbox}[enhanced,breakable,colback=trajredbg,colframe=trajred!45,boxrule=0.35pt,arc=1mm,left=2mm,right=2mm,top=1mm,bottom=1mm,before skip=2mm,after skip=0mm]
{\small \textbf{dev-20260803-184604}\quad \textbf{Terminal 0.0}\quad \textbf{94 steps}\quad \textbf{81 Bash calls}\quad \textbf{wrote /app/gpt2.c}}
\end{tcolorbox}
\end{trajectorybox}

\begin{trajectorybox}
{\sffamily\small\bfseries\color{seedaccent} 11}\hspace{1.2mm}{\sffamily\bfseries Remove unused seed remnants and immediately rerun the terminal path}\par
\vspace{0.6mm}{\sffamily\scriptsize\color{trajmuted} Source closeout and local invocation 4/5}\par
\vspace{1.8mm}
{\sffamily\small\bfseries\color{trajblue} Observation.} Reference searches showed that the new control flow no longer used primitives/ or io\_contract.py. llm\_client.py remained a real fallback when runtime\_llm import failed.
\vspace{1.5mm}
{\sffamily\small\bfseries\color{trajpurple} Modification.} Delete the unused primitive modules and io\_contract.py, then rerun imports and a terminal task.
\begin{diffbox}
\noindent\texttt{- harness/primitives/\{files,git\_ops,paths,process,search\}.py}\par
\noindent\texttt{- harness/io\_contract.py}\par
\noindent\texttt{+ retain harness/llm\_client.py as model.py fallback}\par
\end{diffbox}
{\sffamily\small\bfseries\color{trajgreen} Evaluation.} The post-prune smoke created out.txt containing \texttt{42}; result.json returned \texttt{status=success}.
\end{trajectorybox}

\begin{trajectorybox}
{\sffamily\small\bfseries\color{seedaccent} 12}\hspace{1.2mm}{\sffamily\bfseries Verify the task-json CLI form and every required output artifact}\par
\vspace{0.6mm}{\sffamily\scriptsize\color{trajmuted} Local invocation 5/5}\par
\vspace{1.8mm}
{\sffamily\small\bfseries\color{trajgreen} Evaluation.} Using \texttt{--task-json}, \texttt{--workdir}, and \texttt{--output-dir}, the harness created ok.txt containing \texttt{DONE}. The output directory contained changed\_files.json, patch.diff, response.md, result.json, stderr.log, stdout.log, and trajectory.jsonl.
\vspace{1.5mm}
{\sffamily\small\bfseries\color{trajamber} Decision.} The pruned final package still satisfied the alternate invocation contract and complete artifact contract.
\end{trajectorybox}

\begin{trajectorybox}
{\sffamily\small\bfseries\color{seedaccent} 13}\hspace{1.2mm}{\sffamily\bfseries Close with verified interfaces but no benchmark-driven harness revision}\par
\vspace{0.6mm}{\sffamily\scriptsize\color{trajmuted} Final decision}\par
\vspace{1.8mm}
{\sffamily\small\bfseries\color{trajblue} Observation.} Opus checked all three CLI forms, target-file contents, seven required artifact classes, module imports, and the package listing. The final harness/*.py contained 1,876 lines.
\vspace{1.5mm}
{\sffamily\small\bfseries\color{trajamber} Final decision.} Record contract smoke=1, SWE=0, and Terminal=0. The run demonstrates broad up-front system construction and execution coverage, but not a core harness edit caused by either settled zero.
\end{trajectorybox}

\section{RQ2 Code: GPT-5.5}
\begin{tcbraster}[raster columns=4,raster equal height=rows,raster column skip=1.5mm,raster row skip=2mm]
\begin{metricbox}
\metricvalue{51.0 / 67.416}
\metriclabel{H0 \textperiodcentered{} SWE / Terminal}
\end{metricbox}
\begin{metricbox}
\metricvalue{T2 \textperiodcentered{} 56.0 / 74.157}
\metriclabel{final selection}
\end{metricbox}
\begin{metricbox}
\metricvalue{+5.87 pp}
\metriclabel{pair gain}
\end{metricbox}
\begin{metricbox}
\metricvalue{7}
\metriclabel{complete evaluation loops}
\end{metricbox}
\end{tcbraster}
\vspace{1.5mm}
\begin{trajectorybox}
{\sffamily\small\bfseries\color{seedaccent} 01}\hspace{1.2mm}{\sffamily\bfseries Rewrite the architecture before complete H0 was available}\par
\vspace{0.6mm}{\sffamily\scriptsize\color{trajmuted} H0 $\rightarrow$ T1 \textcolor{trajmuted!55}{$\vert$} 0b33583 \textperiodcentered{} core 5038dd0 \textcolor{trajmuted!55}{$\vert$} edit first \textcolor{trajmuted!55}{$\vert$} 4 files \textcolor{trajmuted!55}{$\vert$} +426 / -48}\par
\vspace{1.8mm}
{\sffamily\small\bfseries\color{trajblue} Observation and analysis.} Complete settled H0 was not available. From the source and a local compileall failure, the creator inferred a Python 3 entry point and predicted Terminal failures from implicit artifact paths, permissions, and repeated actions.
\vspace{1.5mm}
{\sffamily\small\bfseries\color{trajpurple} Modification and core diff.} Add stat\_path, mkdir, chmod, and plan tools; artifact-path discovery; JSON-action recovery; repeated-action unlock; and a manifest entry point that prefers python3.
\begin{diffbox}
\noindent\texttt{- basic read/write/command tools and weak artifact state}\par
\noindent\texttt{+ stat\_path / mkdir / chmod / plan}\par
\noindent\texttt{+ artifact-path discovery and JSON-action recovery}\par
\noindent\texttt{+ manifest entry point prefers python3}\par
\end{diffbox}
\begin{itemize}[leftmargin=1.45em,itemsep=0.4mm,topsep=1mm,parsep=0pt]
\item H0 was later confirmed as SWE 51/100 and Terminal 60/89.
\item This round was architecture-first rather than feedback-driven.
\end{itemize}
\begin{tcolorbox}[enhanced,breakable,colback=trajredbg,colframe=trajred!45,boxrule=0.35pt,arc=1mm,left=2mm,right=2mm,top=1mm,bottom=1mm,before skip=2mm,after skip=0mm]
{\small \textbf{SWE 48.0}\quad \textbf{Terminal 65.169}\quad \textbf{vs. H0: SWE -3, Terminal -2.25}}
\end{tcolorbox}
\end{trajectorybox}
\begin{trajectorybox}
{\sffamily\small\bfseries\color{seedaccent} 02}\hspace{1.2mm}{\sffamily\bfseries Replace weak completion checks with a final-review gate}\par
\vspace{0.6mm}{\sffamily\scriptsize\color{trajmuted} T1 $\rightarrow$ T2 \textcolor{trajmuted!55}{$\vert$} 55c501c \textcolor{trajmuted!55}{$\vert$} selected final \textcolor{trajmuted!55}{$\vert$} 1 file \textcolor{trajmuted!55}{$\vert$} +70 / -1}\par
\vspace{1.8mm}
{\sffamily\small\bfseries\color{trajblue} Observation and analysis.} Both T1 benchmarks regressed. Terminal failures showed that the adapter had not crashed; the agent declared success after weak self-written checks. SWE inspection covered only a few cases.
\vspace{1.5mm}
{\sffamily\small\bfseries\color{trajpurple} Modification and core diff.} Block immediate success under low evidence or failed verification; add \_final\_review\_if\_needed(); and check exact paths, permissions, output format, and real verification before finish.
\begin{diffbox}
\noindent\texttt{- low-evidence tasks could finish immediately}\par
\noindent\texttt{+ \_final\_review\_if\_needed()}\par
\noindent\texttt{+ path / permission / format / verification review}\par
\end{diffbox}
\begin{itemize}[leftmargin=1.45em,itemsep=0.4mm,topsep=1mm,parsep=0pt]
\item Hypothesis: low evidence or failed verification must block finish(success).
\item This was the only round with a helper-level local assertion.
\end{itemize}
\begin{tcolorbox}[enhanced,breakable,colback=trajgreenbg,colframe=trajgreen!45,boxrule=0.35pt,arc=1mm,left=2mm,right=2mm,top=1mm,bottom=1mm,before skip=2mm,after skip=0mm]
{\small \textbf{SWE 56.0}\quad \textbf{Terminal 74.157}\quad \textbf{vs. T1: SWE +8, Terminal +8.99}}
\end{tcolorbox}
\end{trajectorybox}
\begin{trajectorybox}
{\sffamily\small\bfseries\color{seedaccent} 03}\hspace{1.2mm}{\sffamily\bfseries Extend final review to every successful Terminal task}\par
\vspace{0.6mm}{\sffamily\scriptsize\color{trajmuted} T2 $\rightarrow$ T3 \textcolor{trajmuted!55}{$\vert$} d1f9384 \textperiodcentered{} includes ccec8ee \textcolor{trajmuted!55}{$\vert$} 1 file \textcolor{trajmuted!55}{$\vert$} +129 / -12}\par
\vspace{1.8mm}
{\sffamily\small\bfseries\color{trajblue} Observation and analysis.} After T2 became the current best, the creator judged the gate's trigger surface too narrow and extended final review from low-evidence tasks to all Terminal successes.
\vspace{1.5mm}
{\sffamily\small\bfseries\color{trajpurple} Modification and core diff.} Review every Terminal success and add \_is\_test\_path, \_verification\_gaps, and path-candidate filtering.
\begin{diffbox}
\noindent\texttt{- review only low-evidence Terminal tasks}\par
\noindent\texttt{+ review every Terminal success}\par
\noindent\texttt{+ path and test filtering}\par
\end{diffbox}
\begin{itemize}[leftmargin=1.45em,itemsep=0.4mm,topsep=1mm,parsep=0pt]
\item Task-path extraction and test-path filtering were tightened together.
\item The creator extrapolated an existing mechanism rather than identifying a new failure bucket.
\end{itemize}
\begin{tcolorbox}[enhanced,breakable,colback=trajredbg,colframe=trajred!45,boxrule=0.35pt,arc=1mm,left=2mm,right=2mm,top=1mm,bottom=1mm,before skip=2mm,after skip=0mm]
{\small \textbf{SWE 52.0}\quad \textbf{Terminal 75.281}\quad \textbf{vs. T2: SWE -4, Terminal +1.12}}
\end{tcolorbox}
\end{trajectorybox}
\begin{trajectorybox}
{\sffamily\small\bfseries\color{seedaccent} 04}\hspace{1.2mm}{\sffamily\bfseries Remove verification-gap enforcement after probe regression}\par
\vspace{0.6mm}{\sffamily\scriptsize\color{trajmuted} T3 $\rightarrow$ T4 \textcolor{trajmuted!55}{$\vert$} 4d2c74d \textcolor{trajmuted!55}{$\vert$} probe-driven \textcolor{trajmuted!55}{$\vert$} 1 file \textcolor{trajmuted!55}{$\vert$} +5 / -12}\par
\vspace{1.8mm}
{\sffamily\small\bfseries\color{trajblue} Observation and analysis.} The creator injected verification gaps into the decision prompt. When the fixed Terminal probe fell from 4/5 to 3/5, it retained path filtering and removed gap enforcement.
\vspace{1.5mm}
{\sffamily\small\bfseries\color{trajpurple} Modification and core diff.} Remove forced verification-gap injection while retaining path-candidate filtering and avoiding second intervention in already corrected tasks.
\begin{diffbox}
\noindent\texttt{- force verification gaps into successful completion}\par
\noindent\texttt{+ retain path-candidate filtering}\par
\noindent\texttt{+ prevent review from degrading corrected tasks}\par
\end{diffbox}
\begin{itemize}[leftmargin=1.45em,itemsep=0.4mm,topsep=1mm,parsep=0pt]
\item The probe explicitly controlled whether to roll back.
\item No relationship was estimated between the n=5 probe and full evaluation.
\end{itemize}
\begin{tcolorbox}[enhanced,breakable,colback=trajamberbg,colframe=trajamber!45,boxrule=0.35pt,arc=1mm,left=2mm,right=2mm,top=1mm,bottom=1mm,before skip=2mm,after skip=0mm]
{\small \textbf{SWE 57.0}\quad \textbf{Terminal 73.034}\quad \textbf{vs. T3: SWE +5, Terminal -2.25}}
\end{tcolorbox}
\end{trajectorybox}
\begin{trajectorybox}
{\sffamily\small\bfseries\color{seedaccent} 05}\hspace{1.2mm}{\sffamily\bfseries Track files created by shell commands}\par
\vspace{0.6mm}{\sffamily\scriptsize\color{trajmuted} T4 $\rightarrow$ T5 \textcolor{trajmuted!55}{$\vert$} 3eaac25 \textperiodcentered{} includes b67d86e \textcolor{trajmuted!55}{$\vert$} 1 file \textcolor{trajmuted!55}{$\vert$} +48 / -68}\par
\vspace{1.8mm}
{\sffamily\small\bfseries\color{trajblue} Observation and analysis.} Terminal trajectories showed that shell commands modified files without write\_file, so the artifact tracker never recorded them. The final gate therefore lacked awareness of real artifact changes.
\vspace{1.5mm}
{\sffamily\small\bfseries\color{trajpurple} Modification and core diff.} Add \_file\_snapshot(); compare files before and after run\_command; record new and changed files.
\begin{diffbox}
\noindent\texttt{- track only explicit write tools}\par
\noindent\texttt{+ \_file\_snapshot()}\par
\noindent\texttt{+ before/after run\_command artifact comparison}\par
\end{diffbox}
\begin{itemize}[leftmargin=1.45em,itemsep=0.4mm,topsep=1mm,parsep=0pt]
\item No new SWE failure case was opened.
\item The iteration remained focused on Terminal artifact handling.
\end{itemize}
\begin{tcolorbox}[enhanced,breakable,colback=trajredbg,colframe=trajred!45,boxrule=0.35pt,arc=1mm,left=2mm,right=2mm,top=1mm,bottom=1mm,before skip=2mm,after skip=0mm]
{\small \textbf{SWE 52.0}\quad \textbf{Terminal 71.910}\quad \textbf{vs. T4: SWE -5, Terminal -1.12}}
\end{tcolorbox}
\end{trajectorybox}
\begin{trajectorybox}
{\sffamily\small\bfseries\color{seedaccent} 06}\hspace{1.2mm}{\sffamily\bfseries Preserve output head and tail, then review asynchronous cancellation}\par
\vspace{0.6mm}{\sffamily\scriptsize\color{trajmuted} T5 $\rightarrow$ T6 \textcolor{trajmuted!55}{$\vert$} 615a735 \textperiodcentered{} includes a81df31 \textcolor{trajmuted!55}{$\vert$} probe-driven \textcolor{trajmuted!55}{$\vert$} 2 files \textcolor{trajmuted!55}{$\vert$} +6 / -34}\par
\vspace{1.8mm}
{\sffamily\small\bfseries\color{trajblue} Observation and analysis.} The creator tried retaining the head and tail of long output. After probes returned SWE 0.6 and Terminal 0.8, it narrowed the remaining failure to cancel-async-tasks.
\vspace{1.5mm}
{\sffamily\small\bfseries\color{trajpurple} Modification and core diff.} Keep output head and tail, and require final review to check asynchronous cancellation and residual tasks.
\begin{diffbox}
\noindent\texttt{- keep only the beginning of long output}\par
\noindent\texttt{+ keep output head and tail}\par
\noindent\texttt{+ review async cancellation and residual tasks}\par
\end{diffbox}
\begin{itemize}[leftmargin=1.45em,itemsep=0.4mm,topsep=1mm,parsep=0pt]
\item The probe served as a crash/regression gate.
\item 56c8380 lacked a settled SWE leg and remained transport diagnostic.
\end{itemize}
\begin{tcolorbox}[enhanced,breakable,colback=trajgreenbg,colframe=trajgreen!45,boxrule=0.35pt,arc=1mm,left=2mm,right=2mm,top=1mm,bottom=1mm,before skip=2mm,after skip=0mm]
{\small \textbf{SWE 56.0}\quad \textbf{Terminal 73.034}\quad \textbf{vs. T5: SWE +4, Terminal +1.12}}
\end{tcolorbox}
\end{trajectorybox}
\begin{trajectorybox}
{\sffamily\small\bfseries\color{seedaccent} 07}\hspace{1.2mm}{\sffamily\bfseries Try and revert large-file offsets, then constrain finish actions}\par
\vspace{0.6mm}{\sffamily\scriptsize\color{trajmuted} T6 $\rightarrow$ T7 \textcolor{trajmuted!55}{$\vert$} 0a183a0 \textperiodcentered{} 88aa80b / revert 137a41c \textcolor{trajmuted!55}{$\vert$} rollback \textcolor{trajmuted!55}{$\vert$} 1 file \textcolor{trajmuted!55}{$\vert$} +8 / -3}\par
\vspace{1.8mm}
{\sffamily\small\bfseries\color{trajblue} Observation and analysis.} The creator suspected large files were read from the wrong position. A probe returned Terminal 1.0 but SWE 0.6, so it reverted offset reading and required finish to be a standalone explicit action. Probes then reached 4/5 and 5/5.
\vspace{1.5mm}
{\sffamily\small\bfseries\color{trajpurple} Modification and core diff.} Fully revert requested-offset reads for large files; retain only the standalone finish requirement.
\begin{diffbox}
\noindent\texttt{- requested-offset large-file read (reverted)}\par
\noindent\texttt{+ finish must be a standalone action}\par
\end{diffbox}
\begin{itemize}[leftmargin=1.45em,itemsep=0.4mm,topsep=1mm,parsep=0pt]
\item No compileall or smoke test ran in this round.
\item High probe scores did not transfer to the full evaluation.
\end{itemize}
\begin{tcolorbox}[enhanced,breakable,colback=trajredbg,colframe=trajred!45,boxrule=0.35pt,arc=1mm,left=2mm,right=2mm,top=1mm,bottom=1mm,before skip=2mm,after skip=0mm]
{\small \textbf{SWE 50.0}\quad \textbf{Terminal 68.539}\quad \textbf{vs. T6: SWE -6, Terminal -4.49}}
\end{tcolorbox}
\end{trajectorybox}
\begin{trajectorybox}
{\sffamily\small\bfseries\color{seedaccent} 08}\hspace{1.2mm}{\sffamily\bfseries Select T2 instead of the latest version}\par
\vspace{0.6mm}{\sffamily\scriptsize\color{trajmuted} Final decision \textcolor{trajmuted!55}{$\vert$} declare-final 55c501c \textcolor{trajmuted!55}{$\vert$} 3 remaining pairs unused}\par
\vspace{1.8mm}
{\sffamily\small\bfseries\color{trajblue} Observation and analysis.} The creator maintained a combined-score ledger. Thirteen seconds after T7 settled, it confirmed that T2 remained the true argmax and did not submit another candidate merely to sample a higher score.
\vspace{1.5mm}
{\sffamily\small\bfseries\color{trajpurple} Modification and core diff.} Do not use T7; select T2 / 55c501c; end evolution.
\begin{diffbox}
\noindent\texttt{- latest T7}\par
\noindent\texttt{+ T2 / 55c501c}\par
\noindent\texttt{+ end evolution}\par
\end{diffbox}
\begin{itemize}[leftmargin=1.45em,itemsep=0.4mm,topsep=1mm,parsep=0pt]
\item Final version: SWE 56.0, Terminal 74.157.
\item The creator never repaired its own 54\% tool-call rejection.
\end{itemize}

\end{trajectorybox}

\section{RQ2 Code: Opus-4.8}
\begin{tcbraster}[raster columns=4,raster equal height=rows,raster column skip=1.5mm,raster row skip=2mm]
\begin{metricbox}
\metricvalue{68.0 / 74.157}
\metriclabel{H0 \textperiodcentered{} SWE / Terminal}
\end{metricbox}
\begin{metricbox}
\metricvalue{T3 \textperiodcentered{} 74.0 / 74.157}
\metriclabel{final selection}
\end{metricbox}
\begin{metricbox}
\metricvalue{+3.00 pp}
\metriclabel{pair gain}
\end{metricbox}
\begin{metricbox}
\metricvalue{3}
\metriclabel{complete evaluation loops}
\end{metricbox}
\end{tcbraster}
\vspace{1.5mm}
\begin{trajectorybox}
{\sffamily\small\bfseries\color{seedaccent} 01}\hspace{1.2mm}{\sffamily\bfseries Build failure categories from 189 cases before a structural repair}\par
\vspace{0.6mm}{\sffamily\scriptsize\color{trajmuted} H0 $\rightarrow$ T1 \textcolor{trajmuted!55}{$\vert$} 9d6f78e \textcolor{trajmuted!55}{$\vert$} diagnosis-first \textcolor{trajmuted!55}{$\vert$} 8 files \textcolor{trajmuted!55}{$\vert$} +304 / -18}\par
\vspace{1.8mm}
{\sffamily\small\bfseries\color{trajblue} Observation and analysis.} The creator found 56 SWE events with score=null and refused to score from the surface event stream. It read cases.json, the feedback index, and raw results to confirm H0=68/100, then classified about 10 of 32 SWE failures as hidden-grader build, compile, or import errors.
\vspace{1.5mm}
{\sffamily\small\bfseries\color{trajpurple} Modification and core diff.} Add a pre-finish verification gate, SIGTERM/SIGALRM always-finalize and emergency finalize, exact interface/output constraints, and stronger reliability guidance.
\begin{diffbox}
\noindent\texttt{- success could finish without execution after an edit}\par
\noindent\texttt{+ pre-finish verification gate}\par
\noindent\texttt{+ always-finalize / emergency finalize}\par
\noindent\texttt{+ exact interface and output constraints}\par
\end{diffbox}
\begin{itemize}[leftmargin=1.45em,itemsep=0.4mm,topsep=1mm,parsep=0pt]
\item Thinking-block replay and prompt caching were checked first; latency was rejected as the main lever.
\item Hypothesis: agent self-tests do not imply hidden-grader interface correctness; finish must follow real execution.
\end{itemize}
\begin{tcolorbox}[enhanced,breakable,colback=trajgreenbg,colframe=trajgreen!45,boxrule=0.35pt,arc=1mm,left=2mm,right=2mm,top=1mm,bottom=1mm,before skip=2mm,after skip=0mm]
{\small \textbf{SWE 73.0}\quad \textbf{Terminal 75.281}\quad \textbf{vs. H0: SWE +5, Terminal +1.12}}
\end{tcolorbox}
\end{trajectorybox}
\begin{trajectorybox}
{\sffamily\small\bfseries\color{seedaccent} 02}\hspace{1.2mm}{\sffamily\bfseries Give the model the real git diff for one self-review}\par
\vspace{0.6mm}{\sffamily\scriptsize\color{trajmuted} T1 $\rightarrow$ T2 \textcolor{trajmuted!55}{$\vert$} 881b01b \textcolor{trajmuted!55}{$\vert$} 1 file \textcolor{trajmuted!55}{$\vert$} +68 / -0}\par
\vspace{1.8mm}
{\sffamily\small\bfseries\color{trajblue} Observation and analysis.} After T1 improved, the creator inspected 728 edits and found only 14 str\_replace\_no\_match events (1.8\%), rejecting the stronger-editor route. It attributed remaining failures to subtle logic errors in multi-file changes.
\vspace{1.5mm}
{\sffamily\small\bfseries\color{trajpurple} Modification and core diff.} Add \_current\_diff() and one mandatory diff-grounded self-review before a repository task ends.
\begin{diffbox}
\noindent\texttt{- finish review relied on model memory}\par
\noindent\texttt{+ \_current\_diff()}\par
\noindent\texttt{+ mandatory diff-grounded self-review}\par
\end{diffbox}
\begin{itemize}[leftmargin=1.45em,itemsep=0.4mm,topsep=1mm,parsep=0pt]
\item The probe was explicitly treated as a noisy n=5 crash check.
\item The hypothesis was written to the ledger before the change.
\end{itemize}
\begin{tcolorbox}[enhanced,breakable,colback=trajamberbg,colframe=trajamber!45,boxrule=0.35pt,arc=1mm,left=2mm,right=2mm,top=1mm,bottom=1mm,before skip=2mm,after skip=0mm]
{\small \textbf{SWE 75.0}\quad \textbf{Terminal 73.034}\quad \textbf{vs. T1: SWE +2, Terminal -2.25}}
\end{tcolorbox}
\end{trajectorybox}
\begin{trajectorybox}
{\sffamily\small\bfseries\color{seedaccent} 03}\hspace{1.2mm}{\sffamily\bfseries Extend reflection to non-git and Terminal tasks}\par
\vspace{0.6mm}{\sffamily\scriptsize\color{trajmuted} T2 $\rightarrow$ T3 \textcolor{trajmuted!55}{$\vert$} 76de0e6 \textcolor{trajmuted!55}{$\vert$} 1 file \textcolor{trajmuted!55}{$\vert$} +88 / -10 \textcolor{trajmuted!55}{$\vert$} selected final}\par
\vspace{1.8mm}
{\sffamily\small\bfseries\color{trajblue} Observation and analysis.} Many Terminal artifacts were created through shell commands, so H2's git-diff reflection never fired. The creator recorded write\_file/str\_replace paths and run\_command state so non-git tasks could inspect recent artifacts before finishing.
\vspace{1.5mm}
{\sffamily\small\bfseries\color{trajpurple} Modification and core diff.} Add \_written\_paths, \_ran\_command, and \_recent\_workdir\_files(); run artifact self-review for Terminal and non-git tasks.
\begin{diffbox}
\noindent\texttt{- reflection only for repo tasks with git diff}\par
\noindent\texttt{+ written paths and command state}\par
\noindent\texttt{+ recent workdir files}\par
\noindent\texttt{+ non-git artifact self-review}\par
\end{diffbox}
\begin{itemize}[leftmargin=1.45em,itemsep=0.4mm,topsep=1mm,parsep=0pt]
\item All 25 T2 SWE failures had already triggered reflection.
\item R3 inspected concrete interface mismatches but still focused on general reflection.
\end{itemize}
\begin{tcolorbox}[enhanced,breakable,colback=trajamberbg,colframe=trajamber!45,boxrule=0.35pt,arc=1mm,left=2mm,right=2mm,top=1mm,bottom=1mm,before skip=2mm,after skip=0mm]
{\small \textbf{SWE 74.0}\quad \textbf{Terminal 74.157}\quad \textbf{vs. T2: SWE -1, Terminal +1.12}}
\end{tcolorbox}
\end{trajectorybox}
\begin{trajectorybox}
{\sffamily\small\bfseries\color{seedaccent} 04}\hspace{1.2mm}{\sffamily\bfseries Attempt process-group termination, then revert it}\par
\vspace{0.6mm}{\sffamily\scriptsize\color{trajmuted} No formal T4 \textcolor{trajmuted!55}{$\vert$} 5faaa15 \textperiodcentered{} LEDGER.md only \textcolor{trajmuted!55}{$\vert$} no full evaluation \textcolor{trajmuted!55}{$\vert$} session fork}\par
\vspace{1.8mm}
{\sffamily\small\bfseries\color{trajblue} Observation and analysis.} The creator computed cross-version unions and intersections: SWE 81/65/19 and Terminal 71/59/18 for union/intersection/never-pass. Reflection caused further edits only about 4\% of the time. It then attempted setsid + killpg for Bash timeouts.
\vspace{1.5mm}
{\sffamily\small\bfseries\color{trajpurple} Modification and core diff.} One session added setsid/killpg while another checked out harness/tools.py and reverted it. Final commit 5faaa15 changed only LEDGER.md; no H4 code commit existed.
\begin{diffbox}
\noindent\texttt{+ session A: setsid / killpg}\par
\noindent\texttt{- session B: revert tools.py}\par
\noindent\texttt{+ final commit changes LEDGER.md only}\par
\end{diffbox}
\begin{itemize}[leftmargin=1.45em,itemsep=0.4mm,topsep=1mm,parsep=0pt]
\item The creator judged it too risky to change Bash behavior for 189 tasks because of one edge case.
\item Two concurrent sessions made opposite edits to the same file.
\end{itemize}

\end{trajectorybox}
\enlargethispage{2\baselineskip}
\begin{trajectorybox}
{\sffamily\small\bfseries\color{seedaccent} 05}\hspace{1.2mm}{\sffamily\bfseries Stop sampling and select T3}\par
\vspace{0.6mm}{\sffamily\scriptsize\color{trajmuted} Final decision \textcolor{trajmuted!55}{$\vert$} declare-final 76de0e6 \textcolor{trajmuted!55}{$\vert$} 3 of 10 pairs used}\par
\vspace{1.8mm}
{\sffamily\small\bfseries\color{trajblue} Observation and analysis.} The creator estimated noise of roughly $\pm$3--4 tasks in one full evaluation and judged another trivial commit to be additional sampling rather than meaningful evolution. It declined to spend seven remaining pairs chasing a higher random peak.
\vspace{1.5mm}
{\sffamily\small\bfseries\color{trajpurple} Modification and core diff.} Do not add candidates or repeat samples; select T3 as the balanced, validated, mechanism-complete tie-break.
\begin{diffbox}
\noindent\texttt{- more candidates or repeated samples}\par
\noindent\texttt{+ select T3 / 76de0e6}\par
\end{diffbox}
\begin{itemize}[leftmargin=1.45em,itemsep=0.4mm,topsep=1mm,parsep=0pt]
\item Selected T3: SWE 74.0, Terminal 74.157.
\item T1 pair was higher by about 0.06 pp, while T1--T3 passed the same total number of tasks.
\end{itemize}

\end{trajectorybox}
\endgroup

\end{document}